\documentclass{aa}

\usepackage{graphicx}
\usepackage{subcaption}
\usepackage{txfonts}
\usepackage{lscape}
\usepackage[dvipsnames]{xcolor}
\begin{document}

   \title{K-band GRAVITY/VLTI interferometry of "extreme" Herbig Be stars. The size-luminosity relation revisited\thanks{The main data reduction process is available at https://github.com/marcosarenal/gravity-data-reduction.}}
   \titlerunning{GRAVITY/VLTI interferometry of "extreme" Herbig Be stars}

   \subtitle{}

   \author{P. Marcos-Arenal \inst{1},
            I. Mendigut\'ia \inst{1},
            E. Koumpia  \inst{2},   
            R. D. Oudmaijer \inst{2},
            M. Vioque \inst{3,4},
            J. Guzm\'an-D\'iaz\inst{1},
            C. Wichittanakom \inst{2,5},
            W.J. de Wit \inst{6},
            B. Montesinos \inst{1},
            \and
            J.D. Ilee \inst{2}
          }
    \authorrunning{P. Marcos-Arenal et al.
    }

   \institute{Centro de Astrobiolog\'ia (CSIC-INTA), Departamento de Astrof\'isica, ESA-ESAC Campus, PO Box 78, 28691 Villanueva de la Cañada, Madrid, Spain. \email{pmarcos@cab.inta-csic.es}
   \and
             School of Physics and Astronomy, EC Stoner Building, University of Leeds, Leeds, LS2 9JT, UK
    \and
             Joint ALMA Observatory, Alonso de Córdova 3107, Vitacura, Santiago 763-0355, Chile
    \and
             National Radio Astronomy Observatory, 520 Edgemont Road, Charlottesville, VA 22903, USA
    \and
    Department of Physics, Faculty of Science and Technology, Thammasat University, Rangsit Campus, Pathum Thani 12120, Thailand
    \and
    European Southern Observatory, Casilla 19001, Santiago 19, Chile
   }

   \date{Received 4 March 2021; accepted 25 May 2021}

% \abstract{}{}{}{}{} 
% 5 {} token are mandatory
 
  \abstract
   {It has been hypothesized that the location of Herbig Ae/Be stars (HAeBes) within the empirical relation between the inner disk radius (r$_{in}$), inferred from K-band interferometry and the stellar luminosity (L$_*$), is related to the presence of the innermost gas, the disk-to-star accretion mechanism, the dust disk properties inferred from the spectral energy distributions (SEDs), or a combination of these effects. However, no general observational confirmation has been provided to date.}
   {This work aims to test whether the previously proposed hypotheses do, in fact, serve as a general explanation for the distribution of HAeBes in the size-luminosity diagram.}
   {GRAVITY/VLTI spectro-interferometric observations at $\sim $2.2 $ \mu$m have been obtained for five HBes representing two extreme cases concerning the presence of innermost gas and accretion modes. V590 Mon, PDS 281, and HD 94509 show no excess in the near-ultraviolet, Balmer region of the spectra ($\Delta$D$_B$), indicative of a negligible amount of inner gas and disk-to-star accretion, whereas DG Cir and HD 141926 show such strong $\Delta$D$_B$ values that cannot be reproduced from magnetospheric accretion, but probably come from the alternative boundary layer mechanism. In turn, the sample includes three Group I and two Group II stars based on the Meeus et al. SED classification scheme. Additional data for these and all HAeBes resolved through K-band interferometry have been compiled from the literature and updated using Gaia EDR3 distances, almost doubling previous samples used to analyze the size-luminosity relation.}
   {We find no general trend linking the presence of gas inside the dust destruction radius or the accretion mechanism with the location of HAeBes in the size-luminosity diagram. Similarly, our data do not support the more recent hypothesis linking such a location and the SED groups. Underlying trends are present and must be taken into account when interpreting the size-luminosity correlation. In particular, it cannot be statistically ruled out that this correlation is affected by dependencies of both L$_*$ and r$_{in}$ on the wide range of distances to the sources. Still, it is argued that the size-luminosity correlation is most likely to be physically relevant in spite of the previous statistical warning concerning dependencies on distance.}
   {Different observational approaches have been used to test the main scenarios proposed to explain the scatter of locations of HAeBes in the size-luminosity diagram. However, none of these scenarios have been confirmed as a fitting general explanation and this issue remains an open question.}


   \maketitle
% 

%-------------------------------------------------------------------
%--------------------------------------------------------------------

\section{Introduction}
\label{sec:intro}

 It is widely accepted that stars of practically all masses are initially surrounded by accretion disks that play a fundamental role during their formation \citep[e.g.,][and references therein]{Beltran2016}. Thus, the understanding of early stellar evolution is directly related to our knowledge of the accretion mechanism transporting material from the disk to the star \citep[see, e.g., the review in][]{Mendigutia2020}. In Classical T Tauri stars (CTTs, $0.1 < M_{*}/M_{\odot} < 2$), the consensus is that accretion occurs through the stellar magnetic field that connects the inner gas disk and the central star, according to the magnetospheric accretion scenario \citep[MA,][]{Uchida1985,Koenigl1991,Shu1994}. Magnetic fields in massive young stellar objects (MYSOs, $M_{*}/M_{\odot} > 10$) are, in principle, much smaller \citep[e.g.,][and references therein]{Maryline2015}, for which their gaseous disks could reach the stellar surface and directly accrete through a hot and dense boundary layer \citep[BL,][]{Lynden1974}. The possible transition between MA and BL probably occurs within the Herbig Ae/Be (HAeBe, mostly with $2 < M_{*}/M_{\odot} < 10$) regime, as suggested by different lines of evidence \citep[see, e.g.,][and references therein]{Ababakr2017,Wichittanakom2020,Mendigutia2020}. In particular, although the flux excess in the near-ultraviolet (nUV) Balmer region of the spectra (the "Balmer excess" $\Delta$D$_B$) of CTTs and most HAeBes is interpreted from gas accreting magnetospherically, several early-type HBes show such large $\Delta$D$_B$ values that could not have been be reproduced from MA shock modelling, and which may accrete through a BL \citep{Mendigutia2011a,Fairlamb2015}.

Overall, HAeBes are the most massive stars that still show an optically visible pre-main sequence evolution, given that MYSOs remain embedded in their natal envelopes until they reach the main sequence. Moreover, HAeBes are relatively bright and close, which also make them ideal targets for optical and near infrared (nIR) interferometry \citep[see, e.g., the review in][and references therein]{Kraus2015}. Such an observational technique is unique in that it gathers spatially resolved information at a very close radial distance from the stellar surface ($\sim$ 1 au), which is relevant for understanding the star-disk interaction and the accretion process. 

Interferometric studies have revealed that for most HAeBes, there is a correlation between the dust inner sizes probed through the spatially resolved nIR continuum emission and the stellar luminosity  \citep{Monnier2002,Monnier2005}. This empirical correlation takes the form of r$_{in}$ $\propto$ L$_{*}^{1/2}$, with significant scatter, which is consistent with optically thin inner disks and dust destruction radii of relatively large grains ($\sim$ 1 $\mu$m), with sublimation temperatures in the range 1000-2000 K. Intriguingly, some high-luminosity, early-type HBes were found to fall below their expected positions in the size-luminosity diagram, given that their inner dust sizes are smaller than inferred from their stellar luminosities. As discussed, for instance, in the reviews by \citet{MillanGabet2007} and \citet{Kraus2015}, the most plausible reason to explain the undersized nIR continuum emission of such HBes is the presence of gas very close to the star. This gas would shield the dust from stellar irradiation, allowing it to survive closer to the star and to emit in the nIR on top of the dust emission, making the sources appear more compact. Indeed, the undersized HBes show inner dust sizes that are more consistent with classical models having optically thick gas that reaches the stellar surface \citep[][and references therein]{MillanGabet2007,Kraus2015}. This is in line with the BL accretion scenario, rather than that of the MA.  

Studies based on spectro-interferometry, which are capable of spatially resolving optical and nIR emission lines such as H$\alpha$ and Br$\gamma$, reveal that the spatial distribution of the hot atomic gas is more compact or more extended than the dust sublimation front (depending on the specific HAeBe), which has been associated to accretion and wind processes \citep[e.g.,][and references therein]{Kraus2008,Mendigutia2017}. In fact, although there are works devoted to specific stars that suggest that their comparatively smaller inner dust radii could be explained by the presence of inner gas emitting in the nIR \citep[e.g., MWC 147 in][]{Hone2019}, no general trend has been reported.

Alternative explanations for the different location of HAeBes in the size-luminosity diagram, based on the dust properties, have also been proposed. \citet{Benisty2010} tentatively excluded gas as the origin of the nIR continuum emission, at least for HD 163296, proposing instead very refractory grains as the only cause. However, they also noted that such grains should somehow survive temperatures well above 2000 K, inside the conventional dust sublimation zone. More recently, \citet{Perraut2019} proposed that the SED shape according to the \citet{Meeus2001} classification scheme could be related to the different inner dust sizes from K-band interferometry. According to the \citet{Meeus2001} classification Group I stars show a rising continuum from the IR to the sub-millimeter region that can be fitted by a power-law component plus a cool black body, whereas Group II sources only need the power-law component. Studies based on high-resolution imaging associate Group I sources with the presence of "transitional" disks with large gaps and cavities \citep{Maaskant2013,Honda2015,Garufi2017}. This possible link between Group I sources and transitional disks led \citet{Perraut2019} to propose the hypothesis that the Group I characteristic sizes in the K-band are generally larger than those in Group II for a given luminosity bin, and a possible trend may be present in their data in the 10–100 L$_{\odot}$ range. Although no definitive answer could be provided in that work, the confirmation of the hypothesis by \citet{Perraut2019} may give an important, alternative clue on the divergent location of HAeBes in the size-luminosity diagram. In fact, the classification of SEDs based on the \citet{Meeus2001} scheme is unrelated to the presence of inner gas based on accretion rates \citep[][GD21 hereafter]{Mendigutia2012,Guzmandiaz2021}.

In this work, we incorporate GRAVITY/VLTI \citep{Abuter2017} K-band interferometric data of five HBes in the analysis of the size-luminosity correlation. The particular properties of these sources, described in following sections, may be helpful for understanding the origin of such a correlation and its possible relation with the presence of inner gas, the accretion mechanism, and the SED properties. Section \ref{sec:obs} describes the observations and data reduction. Section \ref{sec:modelling} presents the analysis of this dataset. The re-assessment and interpretation of the size-luminosity correlation is presented in Sect. \ref{sec:analysis}. Our main conclusions are summarized in Sect. \ref{sec:conclusions}.  
%--------------------------------------------------------------------
%--------------------------------------------------------------------
\section{Sample, observations, and data reduction}
\label{sec:obs}
Our sample of stars was chosen from \citet{Fairlamb2015} to represent two extreme cases within the HBe regime: sources with negligible amounts of accreting gas interior to the dust disk in accordance with their null values of $\Delta$D$_B$ (V590 Mon, PDS 281, and HD 94509), and those with large amounts of hot gas very close to the star based on the presence of strong $\Delta$D$_B$ values that cannot be modeled with MA (DG Cir and HD 141926). In addition, the sample includes the two types of SEDs according to the classification in \citet{Meeus2001} and spans a wide-enough range of stellar luminosities (1 $<$ log L$_*$/L$_{\odot}$ $<$ 5) to essentially cover the whole size-luminosity diagram for HAeBes. Additional specific properties of the stars studied in this work are described in Sect. \ref{sec:modelling}.

Each star was observed twice during one night between the end of 2018 and the beginning of 2019 with GRAVITY in the wavelength region of $\sim$ 2.0--2.4 $\mu$m and high spectral resolution ($\sim 4000$). The four 8.2m Unit Telescopes (UTs) were used for all stars, except for HD 141926, which is bright enough to use the four 1.8m Auxiliary Telescopes (ATs). The single-field mode was used, meaning that the interferometric signals of the targets were recorded on the fringe tracker (FT) and the science channel (SC) detectors. Spatially unresolved calibrators with similar K-magnitudes and spectral types were also observed with the same instrumental configuration and typical CAL-SCI sequences. The seeing was between 0.4" and 0.6" in all cases. Table \ref{table:A1} in Appendix \ref{Appendix:observations} shows the observing log, and Figs. \ref{fig:V590Mon_u_v_coverage} to \ref{fig:HD141926_u_v_coverage} show the uv coverage of the observed sources.

The data were reduced with the GRAVITY data reduction pipeline \citep [][version Esoreflex-2.9.1]{Lapeyrere2014}. The same standard procedures were applied both to the targets and calibrators to carry out dark, flat-field, and bad-pixel correction, as well as to extract the interferometric observables. The squared visibilities of the targets were derived by dividing those reduced observables by the unresolved calibrators, whose spectra were also used to remove telluric lines and for the final wavelength calibration. Figures \ref{fig:V590Mon_3plots} to \ref{fig:HD141926_3plots} in Appendix \ref{Appendix:observations} show the interferometric observables around the Br$\gamma$ line for each target star and baseline: fluxes, squared visibilities, and differential phases. 

Finally, Fig. \ref{fig:closure_phases_Brg} in Appendix \ref{Appendix:observations} shows the closure phases measured for each star and telescope triplet. The average closure phase for all sources is $\sim$ 0$^{\circ}$ regardless of the triplet, with a scatter around that value $<$ 5$^{\circ}$. The only possible exception is V590 Mon, whose average closure phase is $\sim$ 20$^{\circ}$, although the noise has a comparable size. Additional details inferred from the previous observational results are described in the next section.  

%--------------------------------------------------------------------
%--------------------------------------------------------------------
\section{Interferometric results}
\label{sec:modelling}
Although simple Keplerian disks are capable of reproducing Br$\gamma$ spectro-interferometric observations in some HAeBes \citep{Kraus2012b,Ellerbroek2015,Mendigutia2015}, hydrogen emission lines are usually interpreted in terms of magnetically driven accretion or winds \citep[e.g.,][]{Muzerolle2001,Kurosawa2013,Tamb2014,Tamb2016,Tamb2020}. However, our sample includes two stars whose strong nUV excess cannot be modelled from MA, and there is a lack of alternative BL models of line emission \citep[][and references therein]{Mendigutia2020}. Therefore, in the following we will focus on deriving the sizes associated to the nIR continuum adjacent to Br$\gamma$, only providing qualitative information on this spectral line. The specifics for each star in the sample are discussed in the next subsections.

\subsection{\object{V590 Mon}}
\label{sec:V590Mon}
V590 Mon (aka LkHa 25, Walker 90) is a HBe star with a stellar temperature of $\sim$ 12500 K and mass of $\sim$ 3 M$_{\odot}$ \citep{Fairlamb2015,Moura2020}. A distance of 689$^{+57}_{-49}$ pc is inferred from the recent Gaia EDR3 parallax \citep{Lindegren2020b}. After applying the zero-correction described in that work, such a distance has been derived simply by inverting the parallax, given that the relative error is $<$ 0.1 \citep{Bailer2020}. No excess in the nUV, Balmer region of the spectrum ($\Delta$D$_{B}$ = 0 magnitudes) was reported by \citet{Fairlamb2015}, for which V590 Mon can be considered as virtually non-accreting based on this direct accretion probe. Although V590 Mon has been considered an accreting source in the past, this could be associated with the significant variability of this star on timescales of decades \citep[][and references therein]{Perez2008,Moura2020}. Indeed, the more recent work by \citet{Moura2020} shows that the H$\alpha$ emission currently shown by V590 Mon is associated to outflowing material in a disk wind and that accretion has a negligible contribution to its H$\alpha$ line profile. A companion at $>$ 5$\arcsec$ from the central object and fainter by 6.6 magnitudes in the K-band was reported by \citet{Thomas2007}. At such a relatively large separation, the binary nature of V590 Mon could not be confirmed by \citet{Wheelwright2010} based solely on optical spectro-astrometry. The SED of V590 Mon can be classified as a Group I according to the \citet{Meeus2001} scheme, based on the rising slope at IR wavelengths \citep[Fig. 2 in][]{Moura2020}.

Our GRAVITY data reveal that the single-peaked Br$\gamma$ emission and the adjacent continuum are resolved at all baselines (Fig. \ref{fig:V590Mon_3plots}). However, there is no difference between the visibilities of the line and the continuum, for which the corresponding emitting regions have comparable sizes. The continuum visibilities in the K-band adjacent to the Br$\gamma$ line observed at different baselines were modeled using the \textit{LITpro} software tool \citep{Tallon-Bosc2008}. In particular, we tested parametric geometrical models consisting of a central point source plus an elongated dust distribution with a common center and the shape of a ring, a uniform disk, and a 2D Gaussian disk, given that these are the most common structures normally used to interpret interferometric data of HAeBes. The K-band excess estimate in \citet{Vioque2018}, based on the same stellar parameters from \citet{Fairlamb2015} that we adopt here, leads to weight contributions of the central star and the dust distributions of $\sim$ 10$\%$ and $\sim$ 90$\%$, which were taken as an initial guess for the fitting procedures. The final weights, as well as the position angles, inclinations, and inner dust sizes resulting from the modelling are shown in Table \ref{table:v590mon}. Reduced values of $\chi^2$ normalized by the number of variables in each model (i.e., $\chi^2_{r}$), are used to quantify the goodness of the fits and are also tabulated.

\begin{table}
\centering
\renewcommand\tabcolsep{2pt}
\caption{Dust continuum model results for V590 Mon}
\label{table:v590mon}
\centering
\begin{tabular}{l c c c c c c}
\hline\hline
Model & W$_{point}$ & W$_{dust}$ & r$_{in}$ & i & PA    &$\chi^2_{r}$\\
      & [$\%$]     & [$\%$]     & [mas]      &[$^\circ$] &[$^\circ$]&   \\ 
\hline\hline
Disk$_{gauss}$  & 23$\pm$2 & 77$\pm$8 & 2.99$\pm$0.02 & 36.30$\pm$0.04 & 122$\pm$5 & 13\\
Ring & 27$\pm$3 & 73$\pm$9 & $\sim$1.67 & 49.75$\pm$0.06 & 68.1$\pm$0.7 & 17\\
Disk$_{unif}$ & 28$\pm$4 & 72$\pm$9 & 4.74$\pm$0.04 & $\sim$0 & $\sim$171 & 18\\
\hline
\hline
\end{tabular}
\begin{minipage}{9cm}

\textbf{Notes.} All models consider a point source plus the dust brightness distribution indicated, both centered at the origin. The final values for the corresponding weights, inner dust radius, inclination angle ($i$, 90$^{\circ}$ for edge-on and 0$^{\circ}$ for pole on), major axis position angle ($PA$, measured from North to East), and reduced $\chi^2$ value are tabulated. Rough values are listed without errorbars when these are comparable or larger than the previous ones.
\end{minipage}
\end{table}

\begin{figure}
   \centering
   \includegraphics[width=8cm]{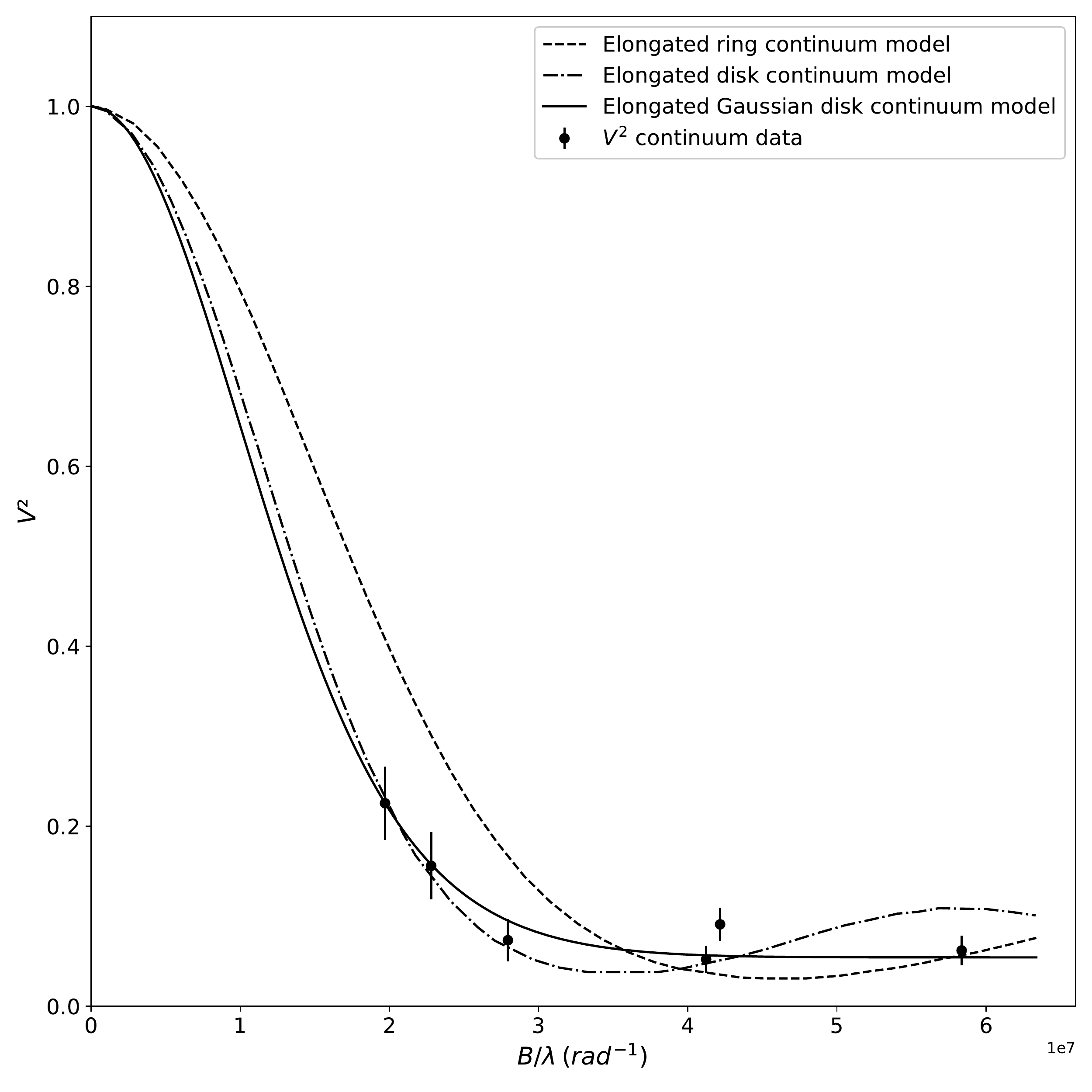}
      \caption{Squared visibilities measured for the continuum adjacent to Br$\gamma$ in V590 Mon as a function of the spatial frequency (solid dots with errorbars). The best fit is indicated with a solid line and corresponds to a Gaussian disk on top of the central star. Two additional fits corresponding to ring and uniform disk models are plotted with dashed and dot-dashed lines. 
      }
         \label{fig:V590Mon_cont_egauss_punt_V2}
   \end{figure}

The Gaussian disk provides the best fit in terms of a smaller $\chi^2_{r}$ value, followed by the ring and the uniform disk. However, all of them result in the same point source contribution within errorbars, larger than the stellar contribution suggested by the K-band excess reported in \citet{Vioque2018}. On top of a possible contribution from unresolved circumstellar emission, this difference may be ascribed to nIR variability or to a stellar temperature different from the one reported in that work (see Sects. \ref{sec:PDS281}, \ref{sec:HD94509}, \ref{sec:DGCir}, and \ref{sec:HD141926}). 
Both the Gaussian disk and the ring models provide an intermediate inclination angle consistent with that resulting from the recent modelling of the H$\alpha$ emission of V590 Mon \citep[40$^{\circ}$ - 55$^{\circ}$;][]{Moura2020}. 
The inclination and major axis position angle ($PA$, measured from North to East) resulting from the uniform disk model have associated errorbars larger than the own values and, as such, the values are unreliable. The PAs inferred from the Gaussian disk and the ring differ significantly, although the dependence on the initial values assumed is negligible in the Gaussian case, but strong in the ring case. The limited uv coverage in our observations prevents us from establishing firm conclusions concerning the PA (see Fig. \ref{fig:V590Mon_u_v_coverage} and Sect. \ref{sec:DGCir}). In the rest of this work, the inner dust size inferred from the Gaussian disk model is assumed, leading to 2.99 $\pm$ 0.02 mas, or 2.1 $\pm$ 0.2 au at the distance to the star. Figure \ref{fig:V590Mon_cont_egauss_punt_V2} shows graphically that such a model is the one that better reproduces the squared visibility data of the nIR continuum adjacent to the Br$\gamma$ emission, as expected from its comparatively smaller $\chi^2_{r}$ value. 

All previously discussed models are centro-symmetric and, as such, they can only reproduce a closure phase close to 0$^{\circ}$. However, V590 Mon shows the largest errorbar associated with the closure phase, which can range from $\sim$ 0$^{\circ}$ to $\sim$ 20$^{\circ}$ (Fig. \ref{fig:closure_phases_Brg} and Sect. \ref{sec:obs}). Such a large errorbar also makes difficult to test whether there is a possible variation of the closure phase with the spatial resolution provided by different baselines, a variation expected if there is a real asymmetry of the source. Still, we tested the eventual presence of a binary that could reproduce the possible closure phase signal of $\sim$ 20$^{\circ}$. This companion should be different from the above mentioned in \citet{Thomas2007}, given that the wide separation reported in that work is orders of magnitude larger than the GRAVITY/VLTI field of view (limited to the Airy disk of each individual aperture, i.e., 60 mas for the UTs in K-band). Although different binary configurations (with and without associated circumstellar structures) are capable of reproducing closure phases close to 20$^{\circ}$, they significantly fail to reproduce the distribution of visibilities versus spatial frequencies shown in Fig. \ref{fig:V590Mon_cont_egauss_punt_V2}. Another possibility for breaking the symmetry is to use models with a single star and an associated circumstellar structure not centered at the same position. However, in order to reproduce a closure phase $\sim$ 20$^{\circ}$ models require that the center of such circumstellar structure falls well outside the GRAVITY field of view, which is physically unrealistic. According to \citet{Perraut2019}, closure phases $\sim$ 25$^{\circ}$ or lower can be reproduced from a combination of inclination effects and different dust grain distributions that introduce some azimuthal modulation in the circumstellar profile. In the case of a confirmed closure phase close to 20$^{\circ}$  in future observations of V590 Mon, this may be a reasonable explanation that is consistent with the anomalous circumstellar dust extinction inferred for this star from nUV spectra \citep{Sitko1984}.

%-------------------------------------------------------------
%                                Figures/V590Mon_cont_egauss_punt_V2
%-------------------------------------------------------------
%
%-------------------------------------------------------------

\subsection{\object{PDS 281}}
\label{sec:PDS281}
PDS 281 (aka SAO 220669) is a HBe star with a stellar temperature of $\sim$ 16000 K \citep{Fairlamb2015,Wichittanakom2020}. Its mass is $\sim$ 9M$_{\odot}$ and it is located at a Gaia EDR3 distance of 868 $^{+9}_{-9}$ pc (GD21). Based on the H$\alpha$ line, \citet{Vieira2003} classified PDS 281 as a relatively evolved HAeBe star. Indeed, the H$\alpha$ emission resulting after the subtraction of the photospheric contribution is weak, the Br$\gamma$ line appears fully in absorption, and $\Delta$D$_B$ = 0 magnitudes, indicating negligible disk-to-star accretion \citep{Fairlamb2015,Fairlamb2017}. Still, the forbidden [\ion{O}{I}] line at 6300 \AA\,is in emission \citep{Fairlamb2017}, possibly indicating the presence of winds. \citet{Vioque2018} reported an IR excess where the circumstellar contribution represents around 30$\%$ of the observed flux at the K-band, with such a value based on stellar parameters compiled from \citet{Fairlamb2015}. The more recent estimates of the stellar parameters based on Gaia distances in \citet{Wichittanakom2020} leads to a null IR excess up to $\sim$ 10 $\mu$m according to GD21. Such a "transitional" SED would be more typical of a relatively evolved young star. GD21 also classified the SED of PDS 281 as Group I.

Figure \ref{fig:PDS281_3plots} shows that the Br$\gamma$ absorption line and the adjacent continuum for PDS 281 are unresolved in our GRAVITY visibilities. In addition, the differential and closure phases are $0^{\circ}$ for all baselines (see also Fig. \ref{fig:closure_phases_Brg}), indicating a common photocentre and a symmetric source for both the line and the continuum. The fact that there is no difference between the visibilities of the absorption line and the continuum suggests that the latter may also be photospheric, in agreement with the null excess at the K-band reported in GD21. An upper limit to the corresponding diameter is given by the best spatial resolution achieved, that is, $\lambda$/2B, with $\lambda$ $\sim$2.17 $\mu$m and B the largest projected baseline for PDS 281 during the observations (130.16 m). The resulting upper limit radius is 0.86 mas (half the angular resolution), or 0.74 au (159 R$_{\odot}$) at the distance to the source. Such an upper limit is well above the stellar radius of PDS 281 estimated by \citet{Wichittanakom2020} and GD21 ($\sim$10R$_{\odot}$), which cannot be resolved with the VLTI. Assuming that the K-band continuum emission is actually photospheric, the size of the inner dust emitting region is in principle much larger than 0.74 au and could be determined observationally by using interferometry at $\sim$ 10 $\mu$m, the shortest wavelength where the IR excess is apparent according to GD21.

\subsection{\object{HD 94509}}
\label{sec:HD94509}

HD 94509 has been commonly catalogued as a HAeBe star \citep[e.g.,][]{Vioque2020}, although it has also been considered an evolved B-star with emission lines \citep[e.g.,][and references therein]{Cowley2015}. The stellar mass is $\sim$ 6 M$_{\odot}$, and it is located at a Gaia EDR3 distance of 1609$^{+45}_{-43}$ pc (GD21). The nUV excess in the Balmer region of the spectrum was reported to be null in \citet{Fairlamb2015}, indicating negligible amounts of gas accreting onto the central star. Still, \citet{Fairlamb2017} report relatively strong emission in H$\alpha$, Br$\gamma$, and [\ion{O}{I}]6300. The nIR SED of HD 94509 was considered to be in excess with respect to the underlying photosphere \citep{Vioque2018} based on a stellar characterization compiled from \citet{Fairlamb2015}, but again the recent estimates in \citet{Wichittanakom2020} and GD21 lead to a null excess up to $\sim$ 3 $\mu$m. The decreasing IR slope in the corresponding SED shown in GD21 suggests that HD 94509 may be a Group II star, although photometry at wavelengths longer than 20 $\mu$m is needed to confirm this.     

Based on our GRAVITY visibilities for this object (Fig. \ref{fig:HD94509_3plots}) the continuum emission remains unresolved for all baselines. The closure phase is $\sim$ 0$^{\circ}$ (Fig. \ref{fig:closure_phases_Brg}), for which the source can be considered centro-symmetric. As for PDS 281 above, an upper limit for the radius of the continuum emitting region can be estimated from the longest baseline in our observations (130.16 m) resulting in $<$ 0.86 mas, or $<$ 1.38 au (298R$_{\odot}$) considering the distance to HD 94509. Once again this upper limit is well above the previous determinations of the stellar radius by \citet{Wichittanakom2020} and GD21 (8-9R$_{\odot}$), which could not be resolved with the VLTI if the nIR continuum is actually photospheric.  

Concerning the double-peaked Br$\gamma$ emission, it is also unresolved for most baselines, except for some residual visibility signal in baseline U4U3, perhaps indicating an emitting region of the blue-shifted peak larger than that of the continuum. In addition, there are clear signals in the differential phases measured in baselines U4U2, U4U1, and U3U1, (with some residuals in U2U1 and U3U2 too) indicating that the photocentres of the blue and red peaks of the Br$\gamma$ emission are displaced with respect to each other and with respect to the continuum. A detailed modeling of the Br$\gamma$ emission in this star is beyond the scope of this work. 
\subsection{DG Cir}
\label{sec:DGCir}

DG Cir (aka VdBH 65b, HBC 596) is a $\sim$ 3.4M$_{\odot}$ HBe star at a Gaia EDR3 distance of 861$^{+15}_{-14}$ pc (GD21) embedded in a nebulosity \citep{Tisserand2013}. 
This star is one of the few HAeBes with a Herbig-Haro source associated, presumably caused by substantial mass loss \citep[][and references therein]{Ray1994}. Indeed, accreting gas is also present, based on such a large nUV excess that cannot be modeled from MA \citep[$\Delta$D$_B$ $\sim$ 0.80 magnitudes;][]{Fairlamb2015}. Excess is also apparent at short, nIR wavelengths, with an SED shape belonging to Group I (GD21). 

Both the single-peaked Br$\gamma$ emission and the adjacent continuum are resolved at all baselines, although there is no difference between the visibilities of the line and the continuum (Fig. \ref{fig:DGCir_3plots}), indicating similar sizes of the corresponding emitting regions. Null values of the differential and closure phases indicate common photocentres and centro-symmetric distributions, respectively. The modeling of the continuum was carried out with \textit{LITpro} in a way similar as for V590 Mon above (Sect. \ref{sec:V590Mon}). The K-band excess reported by \citet{Vioque2018} is based on the same stellar temperature adopted here \citep{Wichittanakom2020} and indicates that the central star contributes with $\sim$ 3$\%$ of the total flux at those wavelengths. Such a weight was assumed as a initial guess for the modeling. On top of the ring and Gaussian/uniform disk distributions that have been tested, the fits significantly improve by adding a background, halo contribution, as happens in other HAeBes \citep[see, e.g.,][]{Lazareff2017,Perraut2019}. Table \ref{table:dgcir} shows the results of the different model fits. Although the Gaussian model best fits the observations in terms of a smaller $\chi^2_{r}$ value, all models provide the same values within the errorbars. The final contribution inferred from the GRAVITY data for the point source is significantly larger than what is expected from the K-band excess mentioned above from \citet{Vioque2018}. Apart from unresolved circumstellar emission and potential nIR variability, uncertainties on the stellar characterization assumed may explain the difference. In fact, the photometric stellar characterization recently made by GD21 leads to a stellar contribution in the K-band of $\sim$ 20$\%$ (see the corresponding SED in GD21), which is in agreement with our estimates from the interferometry. 

\begin{table*}
\centering
\caption{Dust continuum model results for DG Cir}
\label{table:dgcir}
\centering
\begin{tabular}{l c c c c c c c}
\hline\hline
Model & W$_{point}$ & W$_{dust}$ & W$_{bckg}$ & r$_{in}$ & i         & PA       &$\chi^2_{r}$\\
      & [$\%$]     & [$\%$]     & [$\%$]     & [mas]      &[$^\circ$] &[$^\circ$]&\\ 
\hline\hline
Disk$_{gauss}$  & $\sim$20 & 70$^{+30}_{-47}$ & 9.8$\pm$0.4 & 0.4$\pm$0.2 & 56.14$\pm$0.04 & 99.9$\pm$0.8 & 1.802\\
Disk$_{unif}$    & $\sim$20 & $\sim$70  & 9.8$\pm$0.4 & 0.7$\pm$0.4 & 57.10$\pm$0.04 & 100.1$\pm$0.7 & 1.816\\
Ring            & $\sim$20 & 70$\pm$16 & 9.8$\pm$0.4 & $\sim$0.3   & 57.10$\pm$0.04 & 100.1$\pm$0.7 & 1.817\\
\hline
\hline
\end{tabular}
\begin{minipage}{18cm}

\textbf{Notes.} All models consider a point source and a uniform background, plus the dust brightness distribution indicated, all centered at origin. The rest of the notes as in Table \ref{table:v590mon}.
\end{minipage}
\end{table*}

On the other hand, \citet{Ray1994} reported that DG Cir has associated the Herbig-Haro optical flow \object{HH 140}. From a visual inspection of the image in that work, knots A, B, C, and D constituting HH 140 extend along the NW-SE direction at a PA $\sim$ 135$^\circ$-160$^\circ$. This PA is displaced $<$ 60$^\circ$ from the one inferred here and, thus, the outflow would not be roughly perpendicular to the disk as one would expect. In contrast, the intrinsic polarization shown by this star has a position angle $\sim$ 40$^\circ$ \citep{Rodrigues2009}. If such a polarization is tracing the disk major axis, then this would be closer to the perpendicular direction from the HH flow, as expected. However, our fits worsen if the PA of the major axis is forced to be in the NE direction and the best fit goes to the largest angle possible within that quadrant (90$^\circ$). The uv coverage in our observations may bias the PA inferred given that we are probing the NW-SE direction (Fig. \ref{fig:DGCir_u_v_coverage}), and additional observations covering the NE-SW direction would be useful to properly constrain that value. In turn, the intermediate inclination angle inferred for DG Cir, closer to edge-on, is consistent with the presence of P Cygni profiles in optical spectra \citep{Allen1978,Vieira2003} and with that source being classified as an UXor-type variable by the AAVSO\footnote{https://www.aavso.org/}. The uv coverage does not affect the estimated K-band inner dust size, and 0.5 $\pm$ 0.2 mas, or 0.4 $\pm$ 0.2 au at the distance to the star, is adopted hereafter as a reliable value in agreement with the models fitted to the observations.

\begin{figure}
   \centering
   \includegraphics[width=8cm]{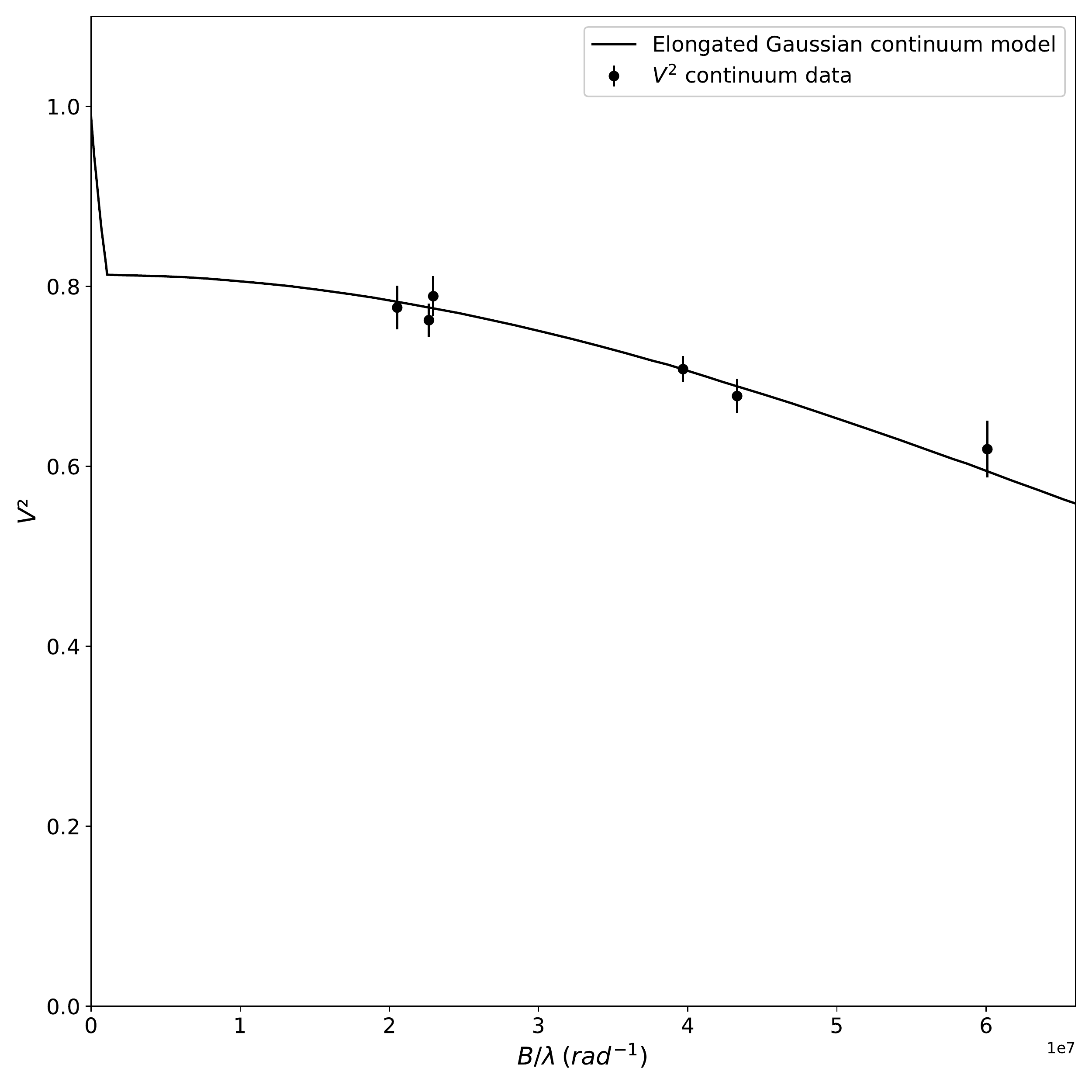}
      \caption{Squared visibilities measured for the continuum adjacent to Br$\gamma$ in DG Cir as a function of the spatial frequency (solid dots with errorbars). The best fit is indicated with a solid line and corresponds to a Gaussian disk on top of the central star, and a background contributing $\sim$ 10$\%$. Two additional fits corresponding to ring and uniform disk models are not plotted as they are practically indistinguishable from the previous. 
      }
         \label{fig:DGCir_cont_egauss_bck_V2}
   \end{figure}

Finally, Fig. \ref{fig:DGCir_cont_egauss_bck_V2} overplots the best fit model in Table \ref{table:dgcir} to the measured squared visibilities versus the spatial frequencies. The rest of the models in that table are not overplotted because there is no visual difference between them.

%-------------------------------------------------------------
%                                Figures/DGCir_cont_egauss_bck_V2
%-------------------------------------------------------------

\subsection{HD 141926}\label{sec:HD141926}
HD 141926 (aka PDS 399, Hen 3-1110, 1H 1555-552) was initially catalogued as a Classical Be star \citep[e.g.,][]{Jaschek1982}, but it was later identified as a HAeBe source \citep{Vieira2003} and considered as such in more recent works (see, e.g., the references in this subsection). HD 141926 is a $\sim$ 17 M$_{\odot}$ HBe star at a Gaia EDR3 distance of $\sim$ 1324$^{+32}_{-31}$ pc (GD21). This source has X-ray emission associated \citep[e.g.,][and references therein]{Maccarone2014,Liu2000}, which adds to the increasing number of HAeBes with such detections \citep[][and references therein]{Stelzer2009}. Concerning the nUV, \citet{Fairlamb2015} found significant excess over the photosphere ($\Delta$D$_B$ $\sim$ 0.20 magnitudes) that could not be reproduced from MA but may be associated with a large amount of accreting gas close to the central star \citep[for a compendium of the emission lines detected in this source see also][]{Fairlamb2017}. An IR excess is also present from short, nIR wavelengths, with the inner dust disk contribution ranging between 30 and 40$\%$ of the total observed flux in the K-band, depending on the stellar parameters adopted (\citeauthor{Vioque2018} 2018; GD21). The overall SED is classified as Group II (GD21). 

Our GRAVITY visibilities reveal that the emitting regions of both the double-peaked Br$\gamma$ line and the adjacent continuum are unresolved (Fig. \ref{fig:HD141926_3plots}). The lack of significant signals in the differential and closure phases (Fig. \ref{fig:closure_phases_Brg}) reveal common photocentres and symmetric distributions for both the emission line and the continuum. An upper limit to the corresponding radius is derived as for PDS 281 and HD 94509 above, but in these cases under the consideration that the largest baseline in our observations was 32.0 m (the ATs were used instead of the UTs, Sect. \ref{sec:obs} and Table \ref{table:A1}). The corresponding dust radius obtained is $<$ 3.49 mas, or $<$ 4.62 au at the distance to the star. 
 
%--------------------------------------------------------------------
%-----------------------------------------------------------------
\section{Analysis of the size-luminosity correlation}\label{sec:analysis}In order to put our previous results in context with the size-luminosity relation, Table \ref{table:comp_stellar} includes relevant data for our stars and for all HAeBes with previous K-band interferometric sizes from the literature (see the references in the caption of that table). Spatially resolved stars discarded as HAeBes in \citet{Vioque2020} have been excluded from the list. The angular inner dust sizes listed in the table were reported in the corresponding references based on model fits of the visibility vs spatial frequency curves, as in our analysis in the previous section. Such angular sizes have been transformed here into spatial sizes using the distances inferred by GD21 based on recent Gaia EDR3 parallaxes \citep[][see also Sect. \ref{sec:V590Mon} for V590 Mon]{Lindegren2020b}. Following GD21, a few distances in the table have been tagged when there is some potential degree of spuriousness, based on uncertainties and flags associated to the new Gaia EDR3 parallaxes. However, as discussed in GD21 the criterion adopted in this work is conservative and the probability of having a correct Gaia EDR3 astrometric solution according to \citet{Rybizki2021} is larger than 0.9 for all stars except for Z CMa A. All distances will be used here, but the reader should be aware of these caveats. Stellar temperatures of most stars were obtained in \citet{Wichittanakom2020} from observed spectra. Adopting the relation with the spectral type in \citet{Kenyon1995}, 9520 K is the effective temperature dividing the sample in hotter HBes and colder HAes. Four stars show temperatures below 7200 K and thus spectral types later than A9. For simplicity, they have been considered as HAes based on previous catalogues (see, e.g., the discussion in GD21) and their relatively large stellar masses ($\geq$ 1.5 M$_{\odot}$). The rest of the stellar parameters (L$_*$, R$_*$, and M$_*$) were mostly derived by GD21 taking those temperatures and the Gaia EDR3 distances as a departure point. The associated errors are mainly based on SED fitting uncertainties, as explained in that work. It is noted that the errors tabulated in Table \ref{table:comp_stellar} are considered symmetric for simplicity, without affecting the analysis in this work. Concerning the distances, almost all sources have small associated errors of the parallaxes ($<$ 1--5$\%$), for which the errors are indeed symmetric for them. We encourage the reader to consult GD21 for more specifics on the distance, the rest of the stellar parameters, and their associated errors. The final sample comprises 48 HAeBes spanning almost 5 orders of magnitude in stellar luminosities (0.5 $<$ log L$_*$/L$_{\odot}$ $<$ 5.2). As such, the following analysis of the K-band size-luminosity relation will be based on the largest sample of HAeBes to date, almost doubling previous samples used for similar analyses \citep{Perraut2019}.

\begin{table*}
\centering
\renewcommand\arraystretch{0.5}
\caption{HAeBes with K-band interferometry. Interferometric sizes and stellar parameters}
\label{table:comp_stellar}
\centering
\begin{tabular}{l l l l l l l l}
\hline\hline
Star        & r$_{in}$                              & d                                          & log r$_{in}$             & log L$_*$         & T$_*$                & R$_*$              & M$_*$             \\
            & [mas]                                 & [pc]                                       & [au]                     & [L$_{\odot}$]     & [K]                  & [R$_ {\odot}$]     & [M$_ {\odot}$]    \\
\hline\hline                                    
V590 Mon    & 2.99$\pm$0.02                         & 689$\pm$53                                 & 0.31$\pm$0.03            & 1.3$\pm$0.4$^a$   & 12500 $\pm$ 1000$^a$ &   2.3$\pm$0.2$^a$  &  3.1$\pm$0.6$^a$ \\
PDS 281     &$<$0.86                                & 868$\pm$9                                  &$<$-0.13                       & 3.72$\pm$0.06     & 16000 $\pm$ 1500     &   9.5$\pm$0.8      &  9.0$\pm$0.5     \\
HD 94509    &$<$0.86                                & 1609$\pm$44                               &$<$0.14                 & 3.11$\pm$0.04     & 11500 $\pm$ 1000     &   8.0$\pm$0.4      &  6.2$\pm$0.2     \\
DG Cir      & 0.50$\pm$0.20                         & 861$\pm$14                                 & -0.37$\pm$0.17           & 2.13$\pm$0.10     & 11000 $\pm$ 3000     &   5.0$\pm$0.6      &  3.4$\pm$0.3     \\
HD 141926   &$<$3.49                                & 1324$\pm$32                                & $<$0.66                       & 4.60$\pm$0.05     & 28000 $\pm$ 1500     &   9.8$\pm$0.7      & 16.9$\pm$0.9     \\
\hline
HD 37806    & (1.8$\pm$0.2)$^1$                     & 397$\pm$4                                  & (-0.15$\pm$0.05)         & 2.30$\pm$0.02     & 10750 $\pm$  156$^b$ &  4.08$\pm$0.16     & 3.50$\pm$0.10     \\
HD 38120    & 3.24$\pm$0.08                         & 381$\pm$5                                  & 0.09$\pm$0.01            & 1.85$\pm$0.02     & 11500 $\pm$  125$^b$ &  2.11$\pm$0.06     & 2.80$\pm$0.04     \\
HD 58647    & (1.7$\pm$0.4)$^2$                     & 302$\pm$2                                  & (-0.29$\pm$0.10)         & 2.49$\pm$0.01     & 10750 $\pm$  125$^b$ &  5.06$\pm$0.14     & 4.05$\pm$0.09     \\
HD 95881    & 1.44$\pm$0.10                         & 1098$\pm$24                                & 0.20$\pm$0.03            & 2.97$\pm$0.02     & 10000 $\pm$  250     &  10.7$\pm$0.4      & 6.40$\pm$0.10     \\
HD 97048    & 2.19$\pm$0.05                         & 184.1$\pm$0.8                              & -0.39$\pm$0.01           & 1.81$\pm$0.01     & 10500 $\pm$  500     &  2.22$\pm$0.06     & 2.80$\pm$0.03     \\
HD 98922    & \emph{2.35$\pm$0.06}$^3$              & 644$\pm$9                                  & \emph{0.18$\pm$0.01}     & 3.16$\pm$0.02     & 10500 $\pm$  250     &  11.4$\pm$0.4      & 7.01$\pm$0.07     \\
HD 100546   & \emph{2.51$\pm$0.06}$^4$              & 108.0$\pm$0.4                              & \emph{-0.57$\pm$0.01}    & 1.34$\pm$0.01     &  9750 $\pm$  500     &  1.83$\pm$0.05     & 2.10$\pm$0.04     \\
HD 135344B  & 1.48$\pm$0.04                         & 134.4$\pm$0.4                              & -0.7$\pm$0.01            & 0.71$\pm$0.01     &  6375 $\pm$  125     &  1.92$\pm$0.08     & 1.46$\pm$0.06     \\
HD 139614   & 4.8$\pm$0.2                           & 133.1$\pm$0.5                              & -0.20$\pm$0.02           & 0.83$\pm$0.01     &  7750 $\pm$  250     &  1.54$\pm$0.05     & 1.60$\pm$0.01     \\
HD 142527   & 1.26$\pm$0.03                         & 158.5$\pm$0.7                              & -0.7$\pm$0.01            & 1.35$\pm$0.01     &  6500 $\pm$  250     &  3.46$\pm$0.13     & 2.20$\pm$0.05     \\
HD 142666   & (1.3$\pm$0.5)$^{2,5}$                 & 145.5$\pm$0.5                              & (-0.74$\pm$0.17)         & 1.13$\pm$0.01     &  7500 $\pm$  250     &  2.33$\pm$0.08     & 1.75$\pm$0.02     \\
HD 144432   & \emph{1.38$\pm$0.04}$^2$              & 154.0$\pm$0.6                              & \emph{-0.67$\pm$0.01}    & 1.21$\pm$0.01     &  7500 $\pm$  250     &  2.23$\pm$0.08     & 1.82$\pm$0.02     \\
HD 144668   & 2.04$\pm$0.05                         & 157.8$\pm$0.8                              & -0.49$\pm$0.01           & 1.97$\pm$0.09     &  8500 $\pm$  250     &   4.7$\pm$0.5      &  3.2$\pm$0.3      \\
HD 145718   & 4.6$\pm$0.3                           & 153.9$\pm$0.5                              & -0.15$\pm$0.03           & 1.08$\pm$0.04     &  8000 $\pm$  250     &  2.05$\pm$0.11     & 1.71$\pm$0.03     \\
HD 150193   & (3$\pm$2)$^{2,6}$                     & 150.0$\pm$0.5                              & (-0.35$\pm$0.29)         & 1.36$\pm$0.01     &  9000 $\pm$  250     &  1.98$\pm$0.06     & 2.25$\pm$0.04     \\
HD 158643   & 3.09$\pm$0.07                         & 125$\pm$2    \textsuperscript{\textdagger} & -0.41$\pm$0.01           & 2.25$\pm$0.01     &  9500 $\pm$  125$^b$ &  4.93$\pm$0.15     & 3.60$\pm$0.05     \\
HD 163296   & (2.2$\pm$1.0)$^{2,3,6,7,8}$           & 100.6$\pm$0.4                              & (-0.65$\pm$0.20)         & 1.19$\pm$0.05     &  9250 $\pm$  250     &  1.70$\pm$0.10     & 1.91$\pm$0.06     \\
HD 169142   & 2.9$\pm$0.6                           & 114.4$\pm$0.4                              & -0.48$\pm$0.10           & 0.76$\pm$0.01     &  7250 $\pm$  125$^b$ &  1.51$\pm$0.05     & 1.55$\pm$0.02     \\
HD 179218   & (12$\pm$10)$^{6,9}$                   & 258$\pm$2                                  & (0.49$\pm$0.36)          & 2.02$\pm$0.01     &  9500 $\pm$  250     &  3.59$\pm$0.10     & 2.99$\pm$0.03     \\
HD 190073   & \emph{2.04$\pm$0.05}$^{1,6,8}$        & 824$\pm$21                                 & \emph{0.23$\pm$0.02}     & 2.88$\pm$0.03     &  9750 $\pm$  250     &   9.7$\pm$0.4      &  6.0$\pm$0.2      \\
HD 259431   & \emph{0.50$\pm$0.02}$^{6,10}$         & 640$\pm$11                                 & \emph{-0.49$\pm$0.02}    & 2.91$\pm$0.06     & 12500 $\pm$  332$^b$ &   6.1$\pm$0.5      &  5.2$\pm$0.3      \\
PDS 27      & 0.83$\pm$0.04                         & 2532$\pm$164 \textsuperscript{\textdagger} & 0.32$\pm$0.03            & 4.00$\pm$0.12     & 17500 $\pm$ 3500     &    15$\pm$2        & 12.0$\pm$1.0      \\
V1818 Ori   & 1.4$\pm$0.3                           & 623$\pm$22   \textsuperscript{\textdagger} & -0.08$\pm$0.11           & 2.67$\pm$0.11     & 11500 $\pm$  125$^b$ &   5.4$\pm$0.7      &  4.5$\pm$0.4      \\
V380 Ori-A  & 1.38$\pm$0.07$^6$                     & 374$\pm$16   \textsuperscript{\textdagger} & -0.29$\pm$0.03           &  2.0$\pm$0.4      & 10250 $\pm$  246$^b$ &   3.1$\pm$1.1      &  2.8$\pm$0.3      \\
MWC 349A    & 10.3$\pm$1.9$^{11}$                   & 1670$\pm$189 \textsuperscript{\textdagger} & 1.23$\pm$0.09            & 5.19$\pm$0.18     & 14000 $\pm$  900$^b$ &    67$\pm$16       &   36$\pm$8        \\
AB Aur      & (2.0$\pm$0.3)$^{1,6}$                 & 155.0$\pm$0.9                              & (-0.51$\pm$0.07)         & 1.66$\pm$0.01     &  9000 $\pm$  125$^b$ &  2.79$\pm$0.09     & 2.36$\pm$0.05     \\
HD 31648    & 1.68$\pm$0.03$^1$                     & 155$\pm$1                                  & -0.58$\pm$0.01           & 1.22$\pm$0.01     &  8000 $\pm$  125$^b$ &  2.13$\pm$0.07     & 1.85$\pm$0.03     \\
CQ Tau      & 1.38$\pm$0.06$^1$                     & 149$\pm$1    \textsuperscript{\textdagger} & -0.69$\pm$0.02           & 0.82$\pm$0.01     &  6750 $\pm$  125$^b$ &  1.88$\pm$0.07     & 1.50$\pm$0.01     \\
T Ori       & \emph{0.82$\pm$0.04}$^{1,6}$          & 399$\pm$4    \textsuperscript{\textdagger} & \emph{-0.49$\pm$0.02}    & 1.77$\pm$0.11     &  9000 $\pm$  500     &   3.0$\pm$0.4      &  2.5$\pm$0.2      \\
MWC 297     &($>$1.68)$^{1,3,6,12,13}$              & 408$\pm$5                                  & ($>$-0.16)                & 4.78$\pm$0.03     & 24000 $\pm$ 2000     &  12.0$\pm$0.6      & 20.0$\pm$0.3      \\
VV Ser      & 1.5$\pm$0.2$^1$                       & 403$\pm$6    \textsuperscript{\textdagger} & -0.23$\pm$0.07           & 2.31$\pm$0.19     & 14000 $\pm$  709$^b$ &   2.4$\pm$0.5      &  3.6$\pm$0.2     \\
MWC 1080    & \emph{1.31$\pm$0.05}$^{1,6}$          & 1424$\pm$62                                & \emph{0.27$\pm$0.03}     & 4.66$\pm$0.13     & 28000 $\pm$ 1735$^b$ &   9.0$\pm$1.7      &   18$\pm$2        \\
UX Ori      & 1.18$\pm$0.16$^2$                     & 320$\pm$3    \textsuperscript{\textdagger} & -0.42$\pm$0.06           & 1.12$\pm$0.17     &  8500 $\pm$  250     &   1.8$\pm$0.3      & 1.91$\pm$0.02     \\
HD 36112    & \emph{1.38$\pm$0.10}$^{1,2}$          & 155.0$\pm$0.8                              & \emph{-0.67$\pm$0.03}    & 0.94$\pm$0.01     &  7250 $\pm$  125$^b$ &  1.87$\pm$0.07     & 1.64$\pm$0.05     \\
Z Cma A     & 1.98$\pm$0.12$^2$                     & 640$\pm$222  \textsuperscript{\textdagger} & 0.10$\pm$0.15            &  3.3$\pm$0.4      &  8250 $\pm$  183$^b$ &    23$\pm$8        &    9$\pm$3        \\
HD 141569   &$<$10.0$^2$                                & 111.1$\pm$0.4                              & $<$0.05                   & 1.40$\pm$0.01     &  9500 $\pm$  250     &  1.75$\pm$0.05     & 2.12$\pm$0.03     \\
HD 143006   & 0.8$\pm$0.5$^2$                       & 166.4$\pm$0.5                              & -0.87$\pm$0.28           & 0.54$\pm$0.02     &  5500 $\pm$  125$^b$ &  2.06$\pm$0.10     & 1.70$\pm$0.10     \\
WW Vul      & 0.90$\pm$0.10$^2$                     & 480$\pm$4                                  & -0.36$\pm$0.05           & 1.41$\pm$0.06     &  8500 $\pm$  125$^b$ &  2.33$\pm$0.16     & 2.04$\pm$0.06     \\
V1685 Cyg   & (1.1$\pm$0.3)$^{1,2,6}$               & 893$\pm$15                                 & (-0.01$\pm$0.12)         & 3.70$\pm$0.02     & 23000 $\pm$ 4000     &   6.5$\pm$0.4      &  8.3$\pm$0.2      \\
V1977 Cyg   & \emph{0.81$\pm$0.08}$^{1,2}$          & 821$\pm$8                                  & \emph{-0.18$\pm$0.04}    & 2.50$\pm$0.01     & 10750 $\pm$  125$^b$ &  5.14$\pm$0.13     & 4.13$\pm$0.09     \\
V1578 Cyg   & 0.78$\pm$0.07$^{2}$                   & 758$\pm$7                                  & -0.23$\pm$0.04           & 2.35$\pm$0.15     & 10500 $\pm$  500     &   4.8$\pm$0.8      &  3.8$\pm$0.5      \\
HD 104237   & 2.48$\pm$0.17$^{3}$                   & 106.5$\pm$0.5                              & -0.58$\pm$0.03           & 1.29$\pm$0.04     &  7750 $\pm$  125$^b$ &  2.44$\pm$0.13     & 1.90$\pm$0.06     \\
V921 Sco    & (2.0$\pm$0.6)$^{3,14}$                & 1399$\pm$69                                & (0.45$\pm$0.13)          & 4.93$\pm$0.12     & 26000 $\pm$  946$^b$ &    14$\pm$2        &   24$\pm$2        \\
                        \hline\hline
\end{tabular}

\begin{minipage}{18.25cm}
\textbf{Notes.} Columns list the name of the star, interferometric inner dust radius, distance based on Gaia EDR3 parallaxes, inner dust radius in spatial scale derived combining r$_{in}$ [mas] and d [pc], stellar luminosity, temperature, radius, and mass. The main references --not indicated with a superscript-- are this work (col 2 for the first five stars), \citet{Perraut2019} (Col. 2 for the following 23 stars up to V1818 Ori), GD21 (cols 3, 5, 7, and 8), and \citet{Wichittanakom2020} (col. 6). Additional references are $^1$\citet{Eisner2004}, $^2$\citet{Monnier2005}, $^3$\citet{Kraus2008}, $^4$\citet{Tatulli2011}, $^5$\citet{Davies2018}, $^6$\citet{MillanGabet2001}, $^7$\citet{Benisty2010}, $^8$\citet{Setterholm2018}, $^9$\citet{Kluska2018},$^{10}$\citet{Hone2019}, $^{11}$\citet{Danchi2001}, $^{12}$\citet{Malbet2007}, $^{13}$\citet{Weigelt2011}, $^{14}$\citet{Kraus2012}, $^a$\citet{Fairlamb2015}, and $^b$GD21. For the sources with more than one value of r$_{in}$ available, the most recent estimate is provided when there are not variations within the errorbars (italics), and an averaged value for the rest (parentheses). Errorbars in the latter cases serve to quantify the ranges provided in the literature. The log L$_*$ value for V590 Mon has been re-scaled from the distance in \citet{Fairlamb2015} to the current Gaia EDR3 distance (Sect. \ref{sec:V590Mon}). Distances tagged with \textsuperscript{\textdagger} may be affected by some degree of spuriousness (see text, and GD21 for details). 
\end{minipage}
\end{table*}

Additional information for the same stars concerning the nUV Balmer excess, the H$\alpha$ and accretion luminosities, the mass accretion rate, the presence of atomic gas inside the dust radius inferred from spectro-interferometry, and the SED group according to the \citet{Meeus2001} classification scheme, is listed in Table \ref{table:comp_others} when possible. This information is further discussed in the following sections.   

\begin{table*}
\centering
\renewcommand\arraystretch{0.5}
\caption{HAeBes with K-band interferometry. Additional parameters}
\label{table:comp_others}
\centering
\begin{tabular}{l l l l l l l}
\hline\hline 

Star        & $\Delta$D$_{B}$   & log L$_{H\alpha}$  & log L$_{acc}$    & log $\dot{M}_{\rm acc}$   & Br$\gamma$/H$\alpha$ &  SED \\
            & [mag]             & [L$_{\odot}$]      & [L$_{\odot}$]    & [M$_{\odot}$ yr$^{-1}$]   & compact?                  & Group\\

\hline\hline  
V590 Mon    & 0.00$\pm$0.05     &  -0.49 $\pm$ 0.07 &   --              &       --                   & no                    & I$^{14}$\\
PDS 281     & 0.00$\pm$0.05     &    ...            &   --              &       --                   & no                    & I\\
HD 94509    & 0.00$\pm$0.05     &   0.29 $\pm$  0.1 &   --              &       --                   & no                    & II?\\
DG Cir      & 0.79$\pm$0.05     &  -0.85 $\pm$ 0.06 &  1.24 $\pm$ 0.07  &     -6.09 $\pm$     0.10   & no                    & I\\
HD 141926   & 0.20$\pm$0.05     &   1.32 $\pm$ 0.09 &  3.41 $\pm$ 0.09  &     -4.32 $\pm$     0.10   & no                    & II\\
\hline               
HD 37806    &...                        &  -0.56 $\pm$ 0.05 &  1.53 $\pm$ 0.07  &     -5.91 $\pm$     0.07   & yes$^1$                & II\\
HD 38120    &...                        &   -0.8 $\pm$ 0.06 &  1.29 $\pm$ 0.07  &     -6.33 $\pm$     0.07   &...                     & I\\
HD 58647    & 0.18$\pm$0.07$^a$ &  -0.43 $\pm$ 0.02 &  1.66 $\pm$ 0.06  &     -5.73 $\pm$     0.07   & yes$^2$                & II\\
HD 95881    & $<$0.05           &   0.33 $\pm$ 0.08 &  2.42 $\pm$ 0.06  &     -4.85 $\pm$     0.06   &...                     & II\\
HD 97048    & $<$0.01           &   -0.9 $\pm$ 0.03 &  1.19 $\pm$ 0.07  &     -6.40 $\pm$     0.08   &...                     & I\\
HD 98922    & $<$0.01           &   0.41 $\pm$ 0.05 &   2.5 $\pm$ 0.06  &     -4.78 $\pm$     0.06   & yes$^{3,4}$           & II\\
HD 100546   & 0.18$\pm$0.05     &   -1.1 $\pm$ 0.03 &  0.99 $\pm$ 0.08  &     -6.57 $\pm$     0.08   & yes$^5$                & I\\
HD 135344B  & 0.07$\pm$0.05     &  -1.99 $\pm$ 0.06 &   0.1 $\pm$ 0.12  &     -7.28 $\pm$     0.12   &...                     & I\\
HD 139614   & 0.09$\pm$0.05     &  -2.04 $\pm$ 0.06 &  0.05 $\pm$ 0.12  &     -7.46 $\pm$     0.12   &...                     & I\\
HD 142527   & 0.06$\pm$0.05     &  -1.45 $\pm$ 0.05 &  0.64 $\pm$ 0.09  &     -6.66 $\pm$     0.10   &...                     & I\\
HD 142666   & $<$0.01           &  -2.03 $\pm$  0.1 &  0.06 $\pm$ 0.12  &     -7.31 $\pm$     0.12   &...                     & II\\
HD 144432   & 0.07$\pm$0.05     &   -1.7 $\pm$ 0.06 &  0.39 $\pm$  0.1  &     -7.02 $\pm$     0.11   &...                     & II\\
HD 144668   & 0.20$\pm$0.05     &  -0.74 $\pm$ 0.04 &  1.35 $\pm$ 0.07  &     -5.98 $\pm$     0.09   &...                     & II\\
HD 145718   & $<$0.01           &  -2.14 $\pm$ 0.12 & -0.05 $\pm$ 0.12  &     -7.46 $\pm$     0.12   &...                     & II\\
HD 150193   & 0.07$\pm$0.05     &  -1.25 $\pm$ 0.05 &  0.84 $\pm$ 0.09  &     -6.72 $\pm$     0.09   &...                     & II\\
HD 158643   & 0.00$\pm$0.07$^a$ &   -0.8 $\pm$ 0.06 &  1.29 $\pm$ 0.07  &     -6.07 $\pm$     0.07   &                            & II\\
HD 163296   & 0.07$\pm$0.05     &  -1.46 $\pm$ 0.05 &  0.63 $\pm$ 0.09  &     -6.91 $\pm$     0.10   & yes$^{3,6}$           & II\\
HD 169142   &...                        &   -1.5 $\pm$ 0.01 &  0.59 $\pm$  0.1  &     -6.92 $\pm$     0.10   &...                     & I\\
HD 179218   & 0.02$\pm$0.07$^a$ &  -0.98 $\pm$ 0.02 &  1.11 $\pm$ 0.08  &     -6.31 $\pm$     0.08   & yes$^7$                & I\\
HD 190073   & 0.22$\pm$0.07$^a$ &   0.27 $\pm$ 0.09 &  2.36 $\pm$ 0.06  &     -4.93 $\pm$     0.07   &...                     & II\\
HD 259431   &...                        &   0.44 $\pm$ 0.06 &  2.53 $\pm$ 0.06  &     -4.90 $\pm$     0.08   & no$^{8}$               & I\\
PDS 27      & 0.17$\pm$0.14     &   1.35 $\pm$ 0.14 &  3.44 $\pm$ 0.09  &     -3.96 $\pm$     0.13   &...                     & I\\
V1818 Ori   &...                        &   0.19 $\pm$ 0.04 &  2.28 $\pm$ 0.06  &     -5.14 $\pm$     0.09   &...                     & I\\
V380 Ori-A  & 0.87$\pm$0.05     &  -0.52 $\pm$ 0.04 &  1.57 $\pm$ 0.07  &     -5.88 $\pm$     0.17   &...                     & I?\\
MWC 349A    &...                        &    ...            &   ...             &       ...                  &...                            & II\\
AB Aur      &...                        &  -0.73 $\pm$ 0.05 &  1.36 $\pm$ 0.07  &     -6.06 $\pm$     0.07   & yes$^9$                & I\\
HD 31648    & 0.05$\pm$0.07$^a$ &   -1.2 $\pm$ 0.03 &  0.89 $\pm$ 0.08  &     -6.55 $\pm$     0.09   &...                     & II\\
CQ Tau      & 0.02$\pm$0.07$^a$ &  -1.98 $\pm$ 0.01 &  0.11 $\pm$ 0.12  &     -7.29 $\pm$     0.12   &...                     & I\\
T Ori       & $<$0.05           &  -1.15 $\pm$ 0.06 &  0.94 $\pm$ 0.08  &     -6.48 $\pm$     0.10   &...                     & ...\\
MWC 297     & $<$0.01           &   2.03 $\pm$ 0.04 &  4.12 $\pm$ 0.12  &     -3.60 $\pm$     0.12   & no$^{3,10,11}$        & I\\
VV Ser      & 0.54$\pm$0.07$^a$ &  -0.47 $\pm$ 0.01 &  1.62 $\pm$ 0.06  &     -6.05 $\pm$     0.12   & yes$^{12,13}$         & II\\
MWC 1080    &...                        &    ...            &   ...             &       ...                  &...                            & I\\
UX Ori      & $<$0.04           &  -1.78 $\pm$ 0.08 &  0.31 $\pm$ 0.11  &     -7.21 $\pm$     0.14   &...                     & II\\
HD 36112    &...                        &  -1.65 $\pm$ 0.03 &  0.44 $\pm$  0.1  &     -7.00 $\pm$     0.10   &...                     & I\\
Z Cma A     & 1.08$\pm$0.05     &   1.62 $\pm$  0.3 &  3.71 $\pm$  0.1  &     -3.40 $\pm$     0.23   &...                     & I\\
HD 141569   & 0.05$\pm$0.05     &  -1.74 $\pm$ 0.08 &  0.35 $\pm$ 0.11  &     -7.23 $\pm$     0.11   & yes$^7$                & II\\
HD 143006   &...                        &  -2.41 $\pm$ 0.04 & -0.32 $\pm$ 0.13  &     -7.74 $\pm$     0.14   &...                     & I\\
WW Vul      & 0.08$\pm$0.07$^a$ &   -1.1 $\pm$ 0.03 &  0.99 $\pm$ 0.08  &     -6.45 $\pm$     0.09   &...                     & II\\
V1685 Cyg   &...                        &    0.9 $\pm$ 0.04 &  2.99 $\pm$ 0.07  &     -4.62 $\pm$     0.08   &...                     & I\\
V1977 Cyg   &...                        &  -0.14 $\pm$ 0.05 &  1.95 $\pm$ 0.06  &     -5.45 $\pm$     0.06   &...                     & II\\
V1578 Cyg   &...                        &  -0.35 $\pm$ 0.04 &  1.74 $\pm$ 0.06  &     -5.66 $\pm$     0.11   &...                     & II\\
HD 104237   & 0.17$\pm$0.05     &  -1.07 $\pm$ 0.03 &  1.02 $\pm$ 0.08  &     -6.37 $\pm$     0.08   & yes$^3$                & II\\
V921 Sco    &...                        &   1.88 $\pm$ 0.16 &  3.97 $\pm$ 0.11  &     -3.75 $\pm$     0.13   & yes$^3$                & I\\
                        \hline\hline
\end{tabular}
\begin{minipage}{18.25cm}
\textbf{Notes.} For the same sample as in Table \ref{table:comp_stellar}, cols. 2-7 indicate the nUV Balmer excess, the H$\alpha$ luminosity, the accretion luminosity, the mass accretion rate, the presence of spectro-interferometric evidence indicating a Br$\gamma$ or H$\alpha$ emitting region more compact than that for the dust emission ("yes", "no" otherwise), and the SED group classification according to the \citet{Meeus2001} scheme. The main references --not indicated with a superscript-- are \citet{Fairlamb2015} (col. 2), the H$\alpha$ luminosities from \citet{Wichittanakom2020}, which have been updated using Gaia EDR3 distances (col. 3), this work (cols. 4, 5, see text; and col 6 for the first five stars), and GD21 (col. 7). The rest of the references are  $^{a}$\citet{Mendigutia2011a}, $^1$\citet{Kreplin2018},  $^2$\citet{Kurosawa2016}, $^3$\citet{Kraus2008}, $^4$\citet{Caratti2015}, $^5$\citet{Mendigutia2015}, $^6$\citet{GarciaLopez2015}, $^7$\citet{Mendigutia2017}, $^{8}$\citet{Hone2019}, $^9$\citet{Perraut2016}, $^{10}$\citet{Malbet2007}, $^{11}$\citet{Weigelt2011}, $^{12}$\citet{GarciaLopez2016}, $^{13}$\citet{Tamb2020}, $^{14}$\citet{Moura2020}. "..." indicates that the corresponding data are not available. "--" indicates that accretion is considered negligible (see text).  
\end{minipage}
\end{table*}

\begin{figure*}
   \centering
   \includegraphics[width=18cm]{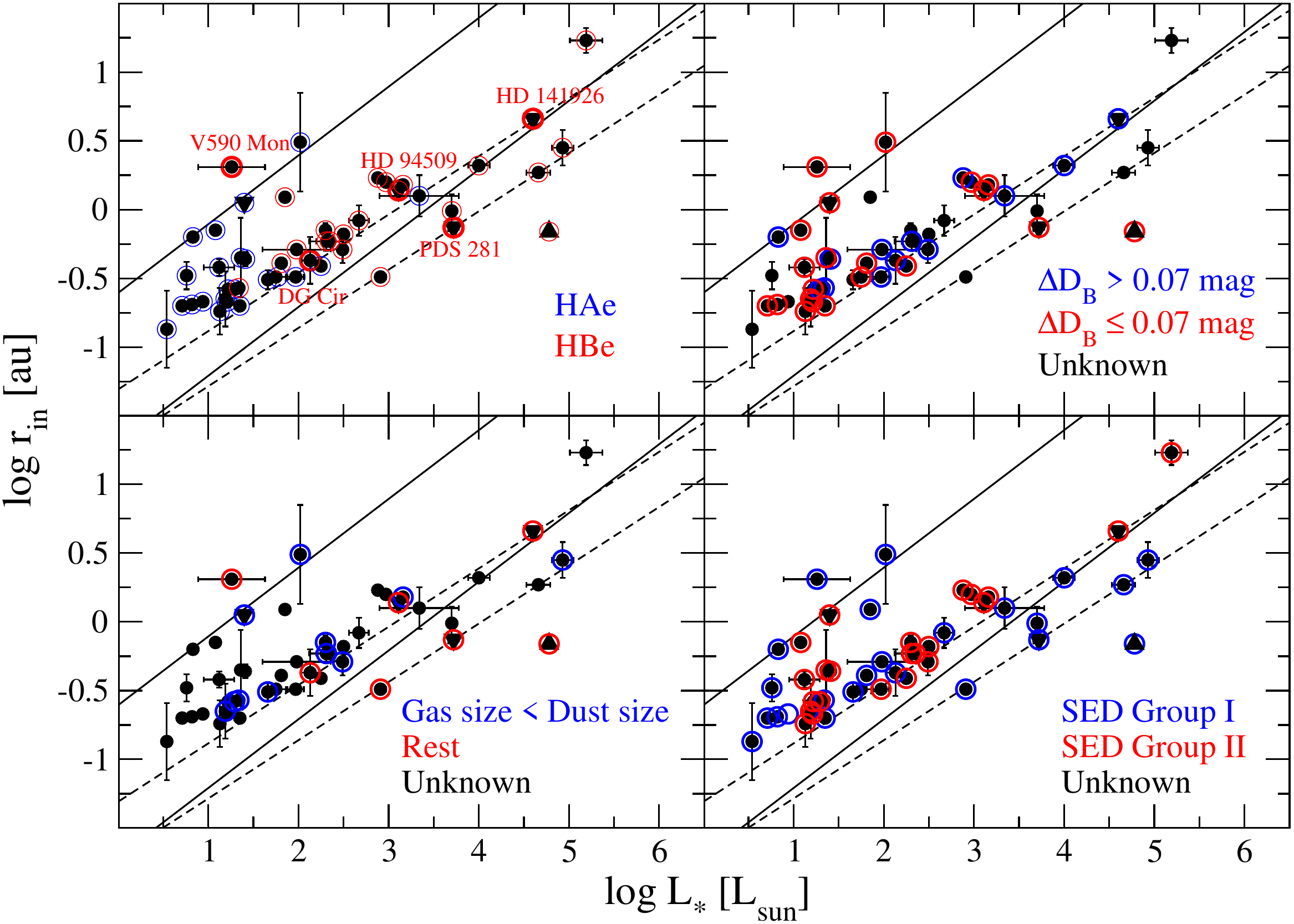}
      \caption{Size-luminosity relation based on data compiled in Tables \ref{table:comp_stellar} and \ref{table:comp_others} for HAeBes. Each panel emphasizes a different aspect, as indicated in the legends. Top left: HAes and HBes; the five with new GRAVITY data presented in this paper are highlighted in red. Top right: The HAeBes with Balmer excess measurements above and below the median value (0.07 magnitudes). Bottom left: The HAeBes with spectro-interferometric measurements indicating that the Br$\gamma$/H$\alpha$ emitting region is smaller than the dust emitting region and the ones without such evidence. Bottom right: Group I and Group II sources based on the SED shape. (All panels): Triangles indicate upper and lower limits, and the solid and dashed lines indicate the expected boundaries of r$_{in}$ for an optically thin-MA and optically thick-BL inner gaseous disks, respectively (see text for details). }
      \label{fig:size_lum}
   \end{figure*}

Figure \ref{fig:size_lum} shows the size-luminosity plot from the data in Table \ref{table:comp_stellar}. The expected inner dust disk sizes corresponding to two different scenarios are indicated with solid and dashed lines. The solid lines refer to the model where the inner disk is optically thin. The stellar luminosity establishes the inner dust radius, located at the region where the sublimation temperature (T$_d$) is reached. In particular, it is assumed that r$_{in}$ (au) = 1.1 $\times$ (Q$_{R}$L$_*$/1000L$_{\odot}$)$^{1/2}$ $\times$ (T$_d$/1500K)$^{-2}$ \citep{Tuthill2001,Monnier2002,Monnier2005}, where Q$_{R}$ = Q$_{abs}$(T$_*$)/Q$_{abs}$(T$_d$) is the ratio of the dust absorption efficiencies Q(T) for radiation at color temperature T of the incident and reemitted field, respectively. The rough boundaries limiting the expected inner dust disk from this model have been overplotted in Fig. \ref{fig:size_lum} with solid lines, obtained assuming values for T$_d$ and Q$_{R}$ of 1000 K and 10 (upper limit), 2000 K and 1 (lower limit). On the other hand, the classical model assumes an optically thick gas disk reaching the central star and located in-between the stellar surface and the inner dust \citep{Hillenbrand1992}. Such a gas disk partially shields the stellar radiation, allowing the dust to survive closer to the star, hence shrinking the value of r$_{in}$. In this case the disk temperature T(r) is assumed to be proportional to r$^{-3/4}$. The proportionality constant depends on the specific stars and is derived from their corresponding R$_*$ and T$_*$ values, thus for each object r$_{in}$ = R$_*$(T$_*$/T$_{d}$)$^{4/3}$. Finally, the log r$_{in}$ values from the previous expression linearly correlate with the corresponding log L$_*$ values for the stars in the sample, and the best linear fits for T$_{d}$ = 1000 K and 2000 K are overplotted in Fig. \ref{fig:size_lum} with dashed lines. These are the boundaries where the dust destruction radius based on the classical model must lie.

It has been remarked that in the classical model, the gas reaches the stellar surface, as in the BL scenario of accretion. Thus, we will use the term "optically thick-BL model" hereafter. In contrast, the above described optically thin model is commonly associated to MA, for which we will refer to it as the "optically thin-MA model". We note, however, that the link in this case is not straightforward, given that in the MA scenario the inner gas may also be optically thick for sources with strong accretion rates \citep{Muzerolle2004}.

According to Fig. \ref{fig:size_lum} (top left), all HAes fall within the region predicted by the optically thin-MA model in the size-luminosity diagram, although the location of 5 out of 23 of such sources ($\sim$ 22$\%$) could also be consistent with the optically thick-BL scenario considering errorbars. In contrast, the distribution of HBes is more heterogeneous. Considering the errorbars, as well as the upper and lower limits, up to 18 and 13 out of 25 of such sources ($<$ 72$\%$ and $<$ 52$\%$) fall within the regions predicted by the optically thin-MA and optically thick-BL models, respectively. This difference between the HAes and many HBes is recovered and expanded here from the initial studies of the size-luminosity relation \citep{Monnier2002,Monnier2005}, as it has been somewhat lost in more recent surveys \citep{Lazareff2017,Perraut2019}. The reason is that these surveys do not include HBes with L$_*$ $>$ 10$^4$L$_{\odot}$, explaining why the previously mentioned difference could not be clearly observed. 

In the following, we test the hypothesis that the presence or absence of gas in between the dust and the star or, otherwise, the different accretion scenarios might explain the locations of HAeBes in the size-luminosity diagram; in particular, the observed scatter $\geq$ $\pm$0.5 dex in r$_{in}$ for a given L$_*$. Other possibilities related with the dust distribution based on the SED shape and with variability is also  considered. Finally, the potential influence of the distances to the sources on the size-luminosity relation is analyzed.

\subsection{Inner dust size and the presence of gas}
\label{sec:gas}
As mentioned before, the main hypothesis for explaining the comparatively smaller r$_{in}$ values shown by some HAeBes (HBes in particular) with respect to others in the size-luminosity diagram is the presence of large amounts of gas in-between the dusty region and the central star, whose shielding effect on the stellar radiation would allow the dust to survive closer to the central source. We note that although predictions from two different models reflecting optically thin and thick inner gaseous disks have been overplotted in Fig. \ref{fig:size_lum}, a smooth transition between the former and the latter scenarios is expected as the mass accretion rate onto the central stars increases \citep[and thus, so does the inner gas density: see][for details]{Muzerolle2004}. In turn, optically thick gaseous disks reaching the central star as in the models by \citet{Hillenbrand1992} are also associated with the BL mechanism of accretion, instead of MA. Different gas probes are used next to test the different aspects of the previous hypothesis.

\subsubsection{Inner gas probed through the Balmer excess}
\label{sec:Balmer}
The main reason for using GRAVITY to observe the five HBes analyzed in Sect. \ref{sec:modelling} is to test the above-mentioned hypothesis based on two types of HBes that it predicts  would be located in very different regions of the size-luminosity diagram. On the one hand, HBes with undetectable Balmer excesses indicating the lack of measurable accretion and thus a negligible amount of gas very close to the central star (V590 Mon, PDS 281, and HD 94509) should have comparatively large r$_{in}$ values. On the other, HBes with large amounts of gas shocking onto the central source as inferred from their large Balmer excesses that cannot be modeled from MA (DG Cir and HD 141926) should have comparatively small r$_{in}$ values. 

The top left panel of Fig. \ref{fig:size_lum} shows the location of the five HBes analyzed in this work in the size-luminosity diagram. Although V590 Mon and HD 94509 show r$_{in}$ values larger than the rest of the stars in similar stellar luminosity bins, as expected from the negligible amount of inner gas inferred from their null values of $\Delta$D$_B$, PDS 281 has a comparatively undersized inner dust size, contrary to the expectations. However, it must be remarked that the K-band sizes derived for PDS 281 and HD 94509 may refer to upper limits of the stellar photospheres and not to the inner dust disks, which could be considerably larger than plotted if the nIR excesses actually start at wavelengths longer than $\sim$ 2 $\mu$m (see the corresponding discussions in Sects. \ref{sec:PDS281} and \ref{sec:HD94509}). On the other hand, DG Cir and HD 141926 do not show significantly smaller r$_{in}$ values compared to HAeBes with similar stellar luminosities, despite the fact that they do not accrete magnetospherically, and the presence of inner gas very close to the star should be large based on their $\Delta$D$_B$ values. These exceptions indicate that the presence or absence of accreting gas closer to the star than the dust or the different accretion mechanisms cannot be the only parameters that determine the distribution of HAeBes in the size-luminosity diagram.
The previous analysis is extended in the top right panel of Fig. \ref{fig:size_lum}  by considering all stars in Table \ref{table:comp_others} with previous measurements of $\Delta$D$_B$ (col. 2). It must be remarked here that $\Delta$D$_B$ is a distance-independent observational quantity from which accretion rates can be directly inferred through appropriate modelling but, regardless of the type of modeling, for a given stellar luminosity bin, large values of $\Delta$D$_B$ may be immediately associated with large accretion rates and the other way around \citep[see][and references therein]{Mendigutia2020}. However, a general trend where the stars with large excesses (indicated in blue in that panel) above the median $\Delta$D$_B$ in our sample (0.07 magnitudes) have comparatively smaller r$_{in}$ values is not observed. The same absence of a trend is obtained when only the ten most extreme cases with the largest and the smallest values of $\Delta$D$_B$ are considered. In addition, apart from DG Cir and HD 141926, VV Ser is the only HAeBe in Table \ref{table:comp_others} whose large excess cannot be modelled from MA \citep{Mendigutia2011a}. However, VV Ser does not show a comparatively undersized inner dust size either. We note that DG Cir, HD 141926, and VV Ser represent $\sim$ 25$\%$ of the HBes in Table \ref{table:comp_others}, with measurements of $\Delta$D$_B$, a percentage similar to the general fraction of HBes that cannot be fitted from MA \citep[based on the stars identified in \citeauthor{Mendigutia2011a} 2011 and \citeauthor{Fairlamb2015} 2015; see][]{Mendigutia2020}. In contrast, the fraction of "undersized" HBes in the size-luminosity diagram is considerably larger than 25$\%$, as described at the beginning of Sect. \ref{sec:analysis}.

In summary, the hypothesis that the presence of a large amount of inner gas or the accretion scenario determines the location of HAeBes in the size-luminosity diagram cannot be confirmed based on measurements of the nUV Balmer excess.

\subsubsection{Inner gas probed through the H$\alpha$ line and accretion}\label{sec:line_and_accretion}

\begin{figure*}
   \centering
   \includegraphics[width=18cm]{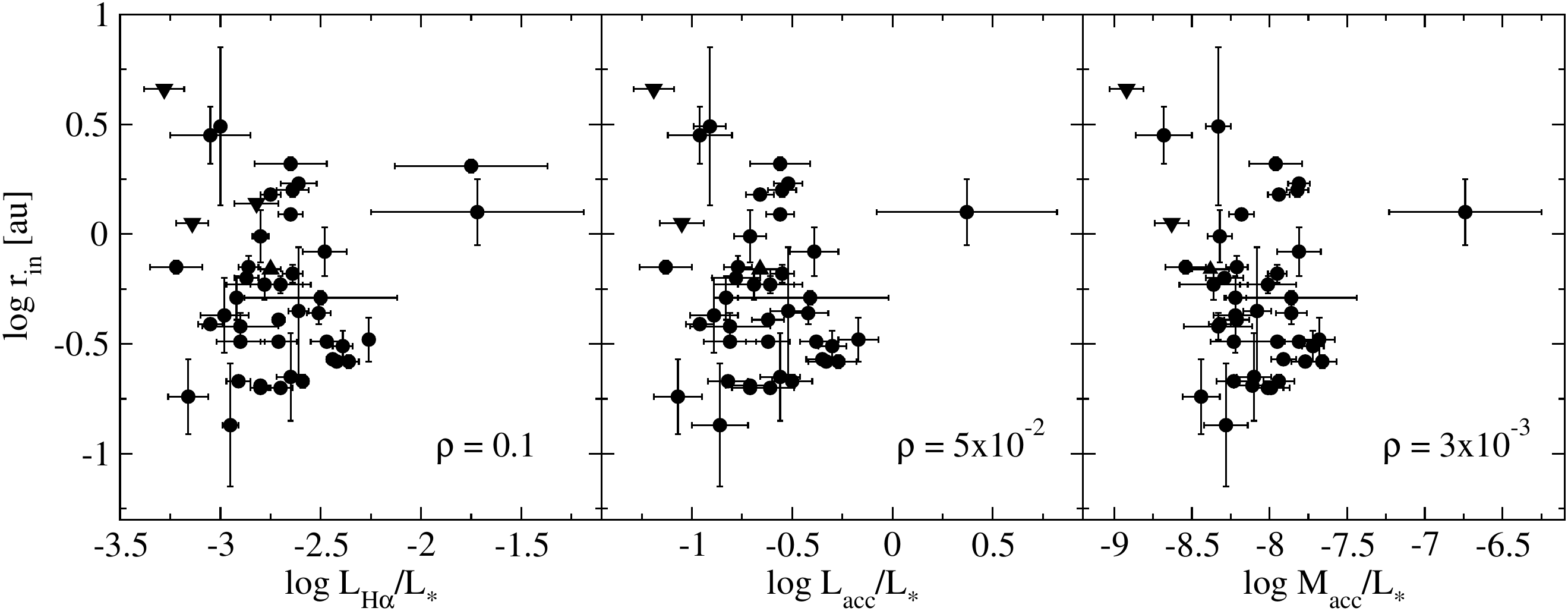}
      \caption{Inner dust size versus the H$\alpha$ luminosity, the accretion luminosity, and the mass accretion rate normalized to the stellar luminosity. Linear correlation coefficients ($\rho$) in each panel quantify the lack of correlation. Upper and lower limits of r$_{in}$ are indicated with triangles.}
      \label{fig:accretion}
   \end{figure*}

Alternatively, the presence of large amounts of gas relatively close to the central star can be inferred from the luminosity of the H$\alpha$ emission line, an observational quantity that is available for almost all stars with K-band continuum interferometry in col. 3 of Table \ref{table:comp_others}. The H$\alpha$ luminosities come from \citet{Wichittanakom2020}, which have been updated using the Gaia EDR3 distances in Table \ref{table:comp_stellar}. In order to make a comparison between log L$_{H\alpha}$ and log r$_{in}$ we must keep in mind the fact that the stellar luminosity correlates with the luminosity of all nUV-optical-nIR emission lines including H$\alpha$ \citep{Mendigutia2015b}, namely, more luminous stars --with larger r$_{in}$ values-- also show stronger H$\alpha$ luminosities. In particular, based on the data in Tables \ref{table:comp_stellar} and \ref{table:comp_others}, the stars in the sample follow a relation L$_{H\alpha}$ $\sim$ L$*$ (linear correlation coefficient $\rho$ = 0.9). Thus, an appropriate comparison should first remove the previous dependence between L$_*$ and L$_{H\alpha}$ dividing the latter by the former. The left panel of Fig. \ref{fig:accretion} compares the inner dust sizes with the ratios between the H$\alpha$ and the stellar luminosities. There is no correlation between the presence of gas directly inferred from the H$\alpha$ luminosities and the inner dust sizes, once L$_{H\alpha}$ is normalized to L$_*$. 

The previous distribution is essentially the same if we use model-dependent accretion luminosities, L$_{acc}$, instead of H$\alpha$ luminosities. The reason is that log L$_{acc}$ and log L$_{H\alpha}$ are related by a simple linear transformation \citep{Mendigutia2011a}. In particular, accretion luminosities in Table \ref{table:comp_others} have been derived from the empirical correlation with the H$\alpha$ luminosity in \citet{Fairlamb2017}\footnote{The H$\alpha$ emission of V590 Mon and HD 94509 has not been converted into accretion luminosities given that both stars are probably non-accreting (Sects. \ref{sec:V590Mon} and \ref{sec:HD94509})}, log (L$_{acc}$/L$_{\odot}$) = 2.09($\pm$0.06) + 1.00($\pm$0.05)$\times$log (L$_{H\alpha}$/L$_{\odot}$). Similarly, given that the mass accretion rate is derived from L$_{acc}$ through the stellar mass and radius ($\dot{M}_{\rm acc}$ = L$_{acc}$R$_*$/GM$_*$), when $\dot{M}_{\rm acc}$ values are inferred either from MA or from BL they also tend to be larger for stars with larger r$_{in}$ values. However, this trend is based on the fact that more massive, that is, more luminous, stars are stronger accretors too \citep[][and references therein]{Mendigutia2015b,Wichittanakom2020}. From the data in Tables \ref{table:comp_stellar} and \ref{table:comp_others},  $\dot{M}_{\rm acc}$ $\sim$ L$_*$ ($\rho$ = 0.9). Thus, once again, the dependence with L$_*$ has to be removed before comparing accretion luminosities and rates with the dust inner sizes. The middle and right panels of Fig. \ref{fig:accretion} show that once normalized to L$_*$ there is no correlation ($\rho$ $<$ 0.5) between the accretion luminosities or mass accretion rates and r$_{in}$, in agreement with our analysis above based on model-independent Balmer excesses and H$\alpha$ emission lines. 

In short, the previous tests do not support that strong accretors generally have comparatively smaller r$_{in}$ values than weak accretors considering similar stellar luminosity ranges, despite the fact that the amount of inner gas and their densities are larger for stronger accretors \citep{Muzerolle2004}.

\subsubsection{Inner gas probed through spectro-interferometry}
\label{sec:specint}
A direct probe of the presence of gas inside the dust emitting region, regardless of whether it is accreting or not, comes from spectro-interferometry. Column 6 in Table \ref{table:comp_others} indicates the comparatively small sub-sample of HAeBes for which there are spectro-interferometric observations in Br$\gamma$ and/or H$\alpha$ from the literature. The bottom left panel of Fig. \ref{fig:size_lum} indicates the location of this sub-sample in the size-luminosity diagram when that is classified in two types of sources. First, the ones for which there is interferometric evidence indicating that atomic gas is emitted in a circumstellar region more compact than that of the dust (blue symbols). Secondly, the rest of the sources for which there is no such evidence (red symbols). Although for the sources in this second group the region where gas is emitted could be equally or more extended than the dust emitting region, the exact situation is unknown for unresolved sources like PDS 281, HD 94509, and HD 141926. There is no general trend and stars with compact emission have both large and small values of r$_{in}$, and the same happens to the sources in the second group. However, the sample of HAeBes with spectro-interferometric measurements is still too scarce to infer general conclusions, with less than 38$\%$ of the whole sample in Table \ref{table:comp_others} having such measurements. 

Additional spectro-interferometric measurements of larger samples of HAeBes, for instance, using the GRAVITY/VLTI and VEGA/CHARA capabilities to spatially and spectrally resolve the Br$\gamma$ and H$\alpha$ lines, are required. These will be of great help to provide an alternative observational test to the hypothesis of the presence of inner gas as a general cause explaining the different locations of HAeBes in the size-luminosity diagram. We note, however, that the objects for which there is interferometric evidence of a gas emitting region more extended than the inner dust (classified above in the second group) may still have gas inside the dust emitting region too. Thus, a careful modeling of the spectro-interferometric observations is necessary to put constraints to the location of the bulk of the gas emitting region relative to the dust.

Finally, the presence of dense gas can also be probed through molecular transitions. In particular, the CO ro-vibrational emission in the wavelength range $\sim$ 2.25--2.45 $\mu$m observed in a few HAeBes is probably coming from the innermost disk regions, based on spectral modelling \citep[see, e.g.,][]{Carr1989,Bik2004,Ilee2014}. However, none of the five stars in our sample show signatures of CO emission in the GRAVITY data, which is in agreement with the low detection rate in HAeBes \citep{Ilee2014,Ilee2018}. In fact, spatially resolved interferometric observations of CO are only available for one HAeBe star to date \citep[see the works on \object{51 Oph} by][]{Tatulli2008,Koutoulaki2020}, which prevents us from making a comparative study.

\subsection{Inner dust size and the SED properties}
\label{sec:dust}
As mentioned in the introduction, the shape of the SED can be associated with the distribution of the circumstellar dust around the central star and eventually to the presence of gaps and holes. The bottom-right panel of Fig. \ref{fig:size_lum} compares the distribution of Group I versus Group II stars in the size-luminosity diagram. Such a classification is available for almost all sources in the sample (last col. in Table \ref{table:comp_others}). Our more complete dataset does not support the hypothesis by \citet{Perraut2019} linking Group I "transitional" stars with larger K-band characteristic sizes at least for the 10-100 L$_{\odot}$ luminosity range. Considering such a range, both Group I and Group II scatter along relatively small and large values of r$_{in}$ values. Moreover, the situation seems to be the opposite to that proposed by \citet{Perraut2019} for L$_*$ $>$ 10$^3$L$_ {\odot}$ (i.e. mainly HBes), where Group I stars appear undersized with respect to Group II sources. However, this conclusion is based on objects that mostly belong to Group I (9 out of 14). In turn, the recent work by GD21 shows that Group I and Group II stars are roughly equally distributed along the HAe and HBe regimes, for which the HBes analyzed here must be biased towards Group I. Additional Group II HBes are thus necessary to test whether there is a distinct location between Group I and Group II stars in the size-luminosity diagram for this regime.

GD21 also classified the SEDs of most HAeBes known to date according to the shortest wavelength where the IR excess is apparent, which is in principle associated to the radial distance from the star where the dust emission originates. They find that while the sources with IR excesses starting at the J or H photometric bands ("Group JH") dominate in the HAe regime, the HBe sources with L$_*$ $>$ 10$^3$L$_ {\odot}$ mostly show IR excess starting at the K photometric band or longer wavelengths ("Group K"). In contrast, the sample of HBes with L$_*$ $>$ 10$^3$L$_ {\odot}$ in Tables \ref{table:comp_stellar} and \ref{table:comp_others} only includes two Group K stars  --PDS 281 and HD 94509, both belonging to our GRAVITY sample. As discussed in Sects. \ref{sec:PDS281} and \ref{sec:HD94509}, the fact that the K-band emission of the latter two stars is unresolved in our GRAVITY observations (as expected for sources with purely photospheric emission at those wavelengths) does not mean that their dust inner sizes are relatively small. On the contrary, stars with IR excesses starting at wavelengths longer than those corresponding to the K-band are expected to have r$_{in}$ values significantly larger than the sources that show nIR excess and are interferometrically resolved at these wavelengths. As a consequence, a robust analysis of the size-luminosity relation should combine continuum interferometry at the K-band and at longer wavelengths \citep[e.g., with MIDI and MATISSE at VLTI,][]{Menu2015,Kobus2019} of a more complete sample of HAeBes belonging to both Group JH and Group K. Such an analysis is likely to add many more HBes with L$_*$ $>$ 10$^3$L$_ {\odot}$ and large r$_{in}$ values. Thus, the apparent trend currently showing undersized HBes is most probably affected by the fact that the current sample includes a comparatively large number of such sources that have an excess starting at short, nIR wavelengths. In other words, a more complete sample of HBes -- mostly with IR excesses starting at longer wavelengths -- may reveal that the undersized sources are mainly exceptions.   

\subsection{Variability of the inner dust size}
\label{sec:var}
More than one estimate of r$_{in}$ is available in the literature for 19 objects in Table \ref{table:comp_stellar}, from which 10 (53$\%$) show differences above errorbars (see cols 2 and 4 in that table and the corresponding caption). To establish whether the cause of the different r$_{in}$ values provided for a given source is actually a temporal change of the inner dust size or is due to other reasons, such as the use of different instrumentation, baselines, and uv coverage, or slight methodological differences, is out of the scope of this work \citep[detailed analysis in, e.g.,][and references therein]{Kobus2020}. 

Instead, our aim is to remark that such differences alone cannot explain the observed $\geq$ $\pm$0.5 dex scatter in r$_{in}$ for a given L$_*$. Indeed, the variations in r$_{in}$ are smaller by a factor 2 ($<$ $\pm$ 0.2 dex changes, considering deviations from the average value) in all stars except for HD 179218 and MWC 297. Multi-epoch K-Band interferometry of HD 179218 (aka MWC 614) provides r$_{in}$ values of 3$\pm$1 mas with the IOTA instrument \citep{MillanGabet2001}, 24$\pm$1 mas with AMBER \citep[combined with additional high-resolution instrumentation; see][]{Kluska2018}, and 8$\pm$1 mas with GRAVITY \citep{Perraut2019}. Thus, the largest variation is of a factor $\sim$ 8 and, considering deviations from the average value, such changes represent $\sim$ $\pm$ 0.4 dex. Concerning MWC 297, this star has been observed multiple times with different interferometers, as indicated in the references provided in Table \ref{table:comp_stellar}. The smallest reported value of r$_{in}$ is 1.68 mas \citep{Eisner2004}. This value is adopted here as a lower limit, considering that r$_{in}$ can indeed reach more than 12 mas \citep{MillanGabet2001} passing through 2-3 mas \citep{Weigelt2011,Kraus2008,Malbet2007}.

In short, apart from a couple of exceptions, so far the data on hand do not support that the observed $\sim$ $\pm$0.5 dex scatter in the size-luminosity relation can be fully explained from differences in the estimates of the r$_{in}$ values --regardless of the cause of such differences. 

\subsection{Considering how distance may affect the size-luminosity correlation}\label{sec:distance}
\citet{Vinkovic2007} showed that the nIR interferometric visibilities of young stars with luminosities below and above 10$^3$L$_{\odot}$ are intrinsically different regardless of the distance, supporting previous studies suggesting that the associated circumstellar sizes differ between these two groups \citep[see Sect. \ref{sec:analysis} and][]{Monnier2002,Monnier2005}. However, \citet{Vinkovic2007} also removed the dependence with the stellar luminosity in their study. Here, we take a critical look at the possible influence of the distance driving the size-luminosity correlation.

Spurious correlations are the ones for which the observed slope, intercept, and scatter are influenced by a common dependence of both parameters compared on a third one. An established way of potentially obtaining a spurious correlation is through arithmetic transformations of both parameters using a common factor \citep{Pearson1897}. Both terms compared in the size-luminosity diagram, L$_*$ and r$_{in}$, are arithmetically derived using the distance to the sources, d. In fact, the derivation of absolute luminosities involves a multiplication by a factor d$^2$, and the transformation from angular to spatial radii also requires multiplying by a factor d. Therefore, our following analysis on the potential sensitivity of the size-luminosity correlation on the distance is justified.

The partial correlation technique is best suited to deal with this type of statistical issue \citep[e.g.,][]{Wall2003}. It measures the degree of association between two variables (L$_*$ and r$_{in}$) assessing the potential dependence on a third one (d) that may be the underlying reason causing the correlation. The first row of Table \ref{table:stats} lists the probability of false correlation (p) and the linear correlation coefficient ($\rho$) quantifying the strength of the size-luminosity correlation for the whole HAeBe sample in Table \ref{table:comp_stellar}. These quantities have been derived both from standard statistics (left columns, i.e., those that do not consider underlying dependencies) and from partial correlations (right columns, i.e., those after removing the effect of potential common dependencies on the distance). Although the p-value is smaller than the typical threshold to establish a correlation ($<$ 0.05) in both cases, the probability of false correlation inferred from partial correlations is orders of magnitude larger than from standard statistics. Similarly, the linear correlation coefficient is above the typical threshold ($>$ 0.5) from standard statistics, but not from partial correlations. The previous results suggest that although the relation between log L$_*$ and log r$_{in}$ is a proper (non-linear) correlation, this is influenced by a common dependence of both parameters compared on the distance to the sources.

\begin{table*}
\centering
\renewcommand\tabcolsep{10.5pt}
\caption{Statistical results}
\label{table:stats}
\centering
\begin{tabular}{c c c c c c}
\hline\hline
Sample & N$_{stars}$ & p & $\rho$ & p$^{pc}$ & $\rho^{pc}$\\
\hline\hline
HAeBes & 43 & 3x10$^{-9}$ & 0.8 & 4x10$^{-3}$ & 0.4\\
HAeBes (d $<$ 250 pc) & 19 & 0.09 & 0.4 & 0.06 & 0.4\\
HAeBes (250 pc $<$ d $<$ 1000 pc) & 19 & 0.2 & 0.4 & 0.2 & 0.3\\
HAeBes (d $>$ 1000 pc) & 5 & 0.1 & 0.9 & 0.1 & 0.9\\
\hline
TTs & 14 & 0.4 & -0.3 & 0.6 & -0.2\\
TTs \& HAeBes (d $<$ 250 pc) & 33 & 9x10$^{-6}$ & 0.7 & 7x10$^{-6}$ & 0.7\\ 
\hline\hline
\end{tabular}
\begin{minipage}{18cm}

\textbf{Notes.} Spearman's probability of false correlation (p) and correlation coefficient ($\rho$) between log L$_*$ [L$_{\odot}$] and log r$_{in}$ [au] for the samples and number of stars indicated in the first two cols. The last two cols are the same quantities derived from partial correlations, i.e., removing the possible common dependence on the distances to the sources. Stars with upper or lower limits for r$_{in}$ have not been considered in the calculations.    
\end{minipage}
\end{table*}

The same statistical tests were then carried out considering sub-samples located at narrower ranges of distances. The corresponding ranges and results are indicated in the following rows of Table \ref{table:stats}. An additional sub-sample of TTs with nIR interferometric sizes compiled by \citet{Pinte2008} has also been considered (rows 5 and 6 in Table \ref{table:stats}). The r$_{in}$ values listed in their Table 2 have been updated using Gaia EDR3 distances, with all TTs being closer than 200 pc. The comparison between the numbers resulting from standard statistics and from partial correlations indicates that although now the distance does not play a role (the values at the left- and at the right-side columns are similar), there is no statistically significant correlation between log L$_*$ and log r$_{in}$ for any of these sub-samples (the typical cut-offs for p and $\rho$ are not reached). The only exception is the sub-sample made by combining the TTs and the HAeBes closer than 250 pc. The comparison between the statistical quantities in the last row of Table \ref{table:stats} shows that when these sources are taken together there is a statistically significant correlation that is not affected by underlying dependencies with the distance. 

\begin{figure}
   \centering
   \includegraphics[width=8.5cm,clip=true]{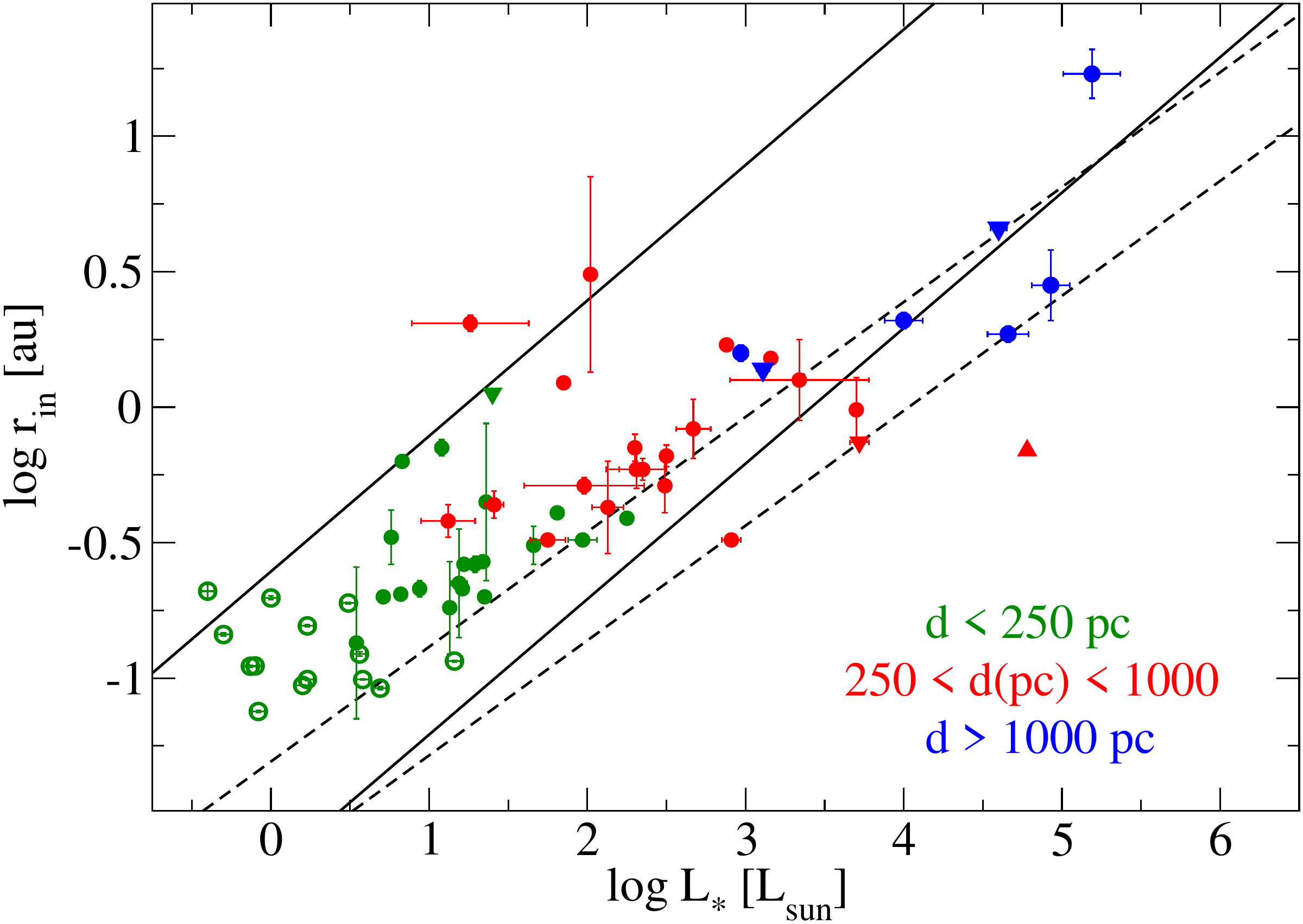}
      \caption{Size-luminosity relation where the stars have been color-coded to indicate different distance ranges as shown in the legend. Interferometrically resolved TTs listed in \citet{Pinte2008} are included in the plot after correction with Gaia EDR3 distances (green open circles),  all of them located between 100 and 200 pc from us. The triangles, solid, and dashed lines represent upper and lower limits and possible trends for the dust destruction radius as in Fig. \ref{fig:size_lum}.       
      }
         \label{fig:size_lum_distance}
   \end{figure}

Figure \ref{fig:size_lum_distance} serves to illustrate the case graphically. This is the same as Fig. \ref{fig:size_lum} but the stars have now been color-coded to indicate the distance ranges where they are located, following the same bins as in Table \ref{table:stats}. Direct visual inspection supports the conclusions based on the statistical analysis included above: from the current sample, it cannot be concluded that there is a significant correlation between log L$_*$ and log r$_{in}$ once sub-samples located in narrow ranges of distance are selected. In fact, the stars in each distance bin show roughly horizontal distributions with similar values of r$_{in}$ within errorbars, despite of the fact that they have stellar luminosities different by orders of magnitude. The only exception is again the sub-sample made of TTs and close HAeBes (green symbols in Fig. \ref{fig:size_lum_distance}). These are visually correlated, mainly falling within the predictions of the optically thin-MA model.  

The previous discussion poses the question about whether the slope, intercept and scatter of the observed size-luminosity correlation for HAeBes carry any physical significance, or if those are fundamentally driven by the distance. In the following, we argue that even when the size-luminosity correlation is indeed affected by underlying dependencies with the distance, the correlation is most likely to be physically meaningful.

\begin{figure}
   \centering
   \includegraphics[width=8cm]{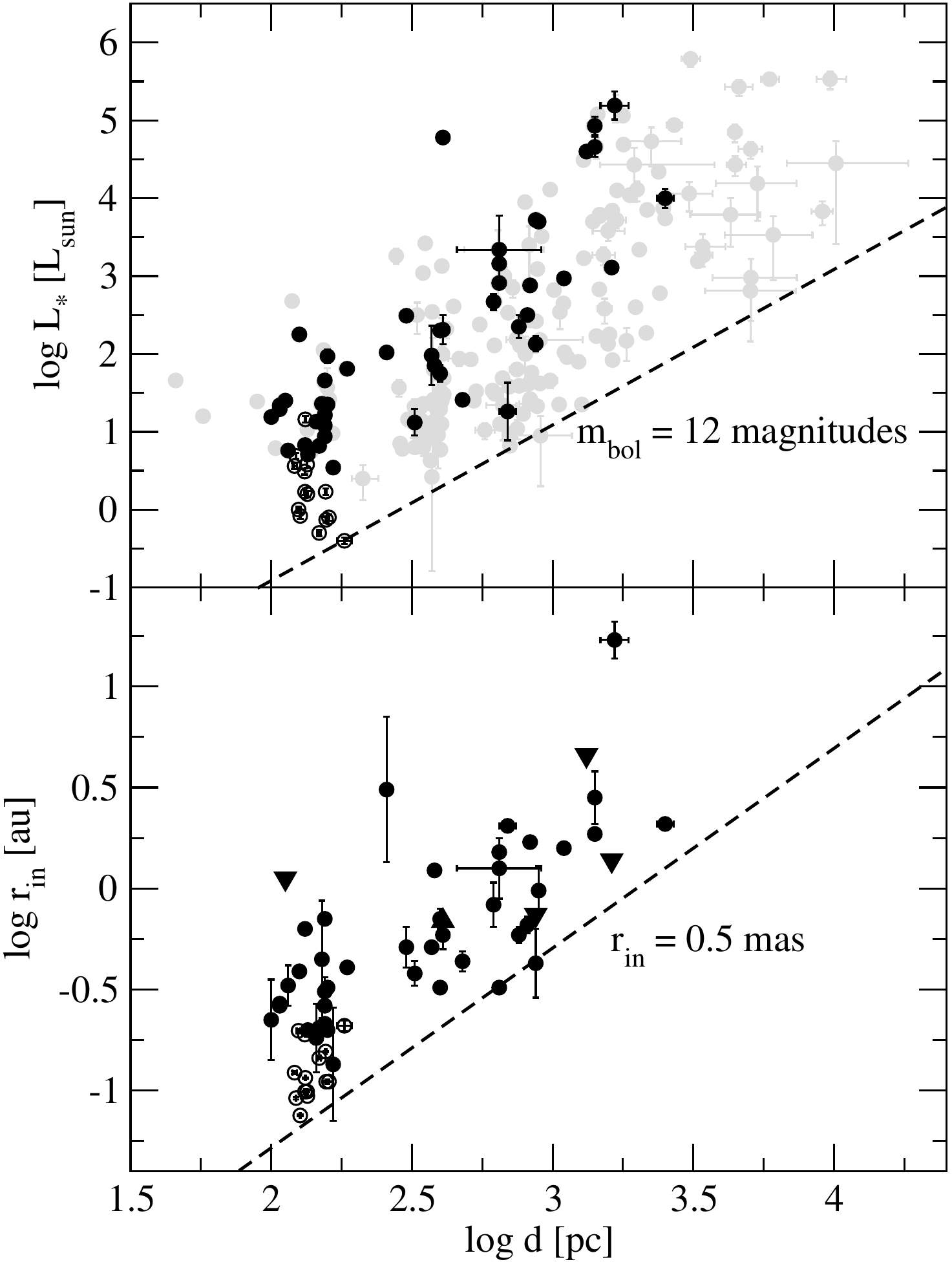}
      \caption{Relation with the distance of the stellar luminosity (top) and the inner dust radius (bottom) for the HAeBes (solid circles and triangles to indicate upper and lower limits), and TTs (open circles) with interferometric estimates of r$_{in}$ discussed in the text. The gray symbols in the top panel represent most HAeBes known to date from the data in GD21. The regions below the dashed lines indicate apparent bolometric magnitudes $>$ 12 and angular radii $<$ 0.5 mas.   
      }
         \label{fig:d_dependence2}
   \end{figure}

The top and bottom panels of Fig. \ref{fig:d_dependence2} show the stellar luminosities and inner dust radii of the sampled HAeBes and TTs versus their Gaia EDR3 distances. For d $<$ 250 pc (log d $<$ 2.4), there is a scatter of points both for L$_*$ and r$_{in}$. Thus, for nearby sources we are probing a relatively broad range of stellar luminosities (10$^{-0.5}$ < L$_*$/L$_{\odot}$ $<$ 10$^{2.5}$) and inner dust radii (0.02 < r$_{in}$/au $<$ 1.8) without any dependence on the distance. Precisely because of the lack of such underlying dependencies, the partial correlation statistics confirmed the result from standard statistics: there is a correlation between L$_*$ and r$_{in}$ for TTs and close HAeBes for which the distance does not play a role. In contrast, from d $>$ 250 pc there is a rising trend linking both L$_*$ and r$_{in}$ with d. These underlying trends explain why the partial correlations statistics warn that the correlation between L$_*$ and r$_{in}$ is affected by the distance. We note that although the arithmetic to calculate both L$_*$ and r$_{in}$ involves the use of d for all sources, it is the presence of underlying correlations between both parameters with d what matters, regardless of the arithmetic. Therefore, in order to assess whether the size-luminosity relation is physically relevant or not, we need first to understand in detail what causes the underlying correlations of both parameters with the distance. 

The trend linking L$_*$ and d is common for most HAeBes known to date, as shown by their corresponding parameters from GD21 (overplotted in gray in the top panel of Fig. \ref{fig:d_dependence2}). The reason explaining such a trend is double. Firstly, high-mass, luminous sources are less frequent and evolve faster than lower-mass stars, for which a larger space volume is necessary to detect them. Secondly, low luminosity sources cannot be observed at long distances due to current observational detection limits. This is indicated by the dashed line, above which only the bright sources with an apparent bolometric magnitude $<$ 12 are located. On the other hand, the dashed line in the bottom panel of Fig. \ref{fig:d_dependence2} roughly represents the smallest angular radius that can be resolved with the VLTI at $\sim$ 2 $\mu$m, $\sim$ 0.5 mas. It is tempting to associate the lower bound of the trend linking r$_{in}$ and d to such angular resolution limits. However, if this were the case, we should see mainly upper limits of r$_{in}$ (i.e., unresolved sources) populating the region near the dashed line, but based on the published data, this is not observed. Instead, the trend linking r$_{in}$ and d seems to be a by-product of the above relation between L$_*$ and d: high luminosity (distant) sources have intrinsically larger inner disk sizes compared to low-luminosity (nearby) sources. Therefore, even though improving our detection and resolution limits would remove the underlying dependencies with the distance, it would only add additional data for distant, low luminosity sources. But low-luminosity stars are already well probed locally, and thus the currently observed size-luminosity correlation is not expected to change significantly. This conclusion involves two assumptions: Firstly, it is assumed that the r$_{in}$ values reported for high luminosity stars are representative. If there is a large amount of unpublished upper limits of r$_{in}$ coinciding with our current resolution limits then the actual size-luminosity diagram differs from the current one. Secondly, it is assumed that the environment and boundary conditions do not have a critical effect on the inner dust radii. If  this was somehow not the case, then the L$_*$--r$_{in}$ correlation could eventually change once distant, low-luminosity stars -- potentially under different conditions than the ones nearby -- are probed. 

In summary, the size-luminosity correlation can be statistically confirmed only for TTs and nearby HAeBes, when both sub-samples are taken together, but the overall correlation is affected by underlying dependencies of both L$_*$ and r$_{in}$ on d. These observed dependencies are caused by a combination of detection limits and the lack of luminous sources in the near vicinity. Improving the detection -- and resolution -- limits in the future will only add additional low-luminosity sources to the size-luminosity diagram. Given that these are already well probed from nearby stars, it is expected that the whole size-luminosity correlation will remain essentially the same in the future. In this sense, and regardless of the statistical warning, the size-luminosity correlation can in principle be interpreted physically despite the underlying dependencies with the distance. However, we  note that the best scenario for studying the size-luminosity relation involves removing the underlying dependencies by considering the widest possible range of stellar luminosities in a narrow enough range of distances. Improving current detection limits with future upgrades such as GRAVITY+\footnote{https://www.mpe.mpg.de/ir/gravityplus} will facilitate such studies. 

\section{Conclusions}\label{sec:conclusions}
This work broadens the sample of HAeBes with K-band spectro-interferometric data by adding five HBes observed with GRAVITY/VLTI. Qualitative information has been provided with respect to the Br$\gamma$ line, but this work has been mainly focused on the analysis of the adjacent continuum. Simple parametrical models have been fitted to the two spatially resolved stars (V590 Mon and DG Cir), providing geometrical estimates of their inner dust distributions. Upper limits for the sizes of the inner dust regions and photospheres are provided for the unresolved stars (PDS 281, HD 94509, and HD 141926).

The previous sample was selected because of the extreme properties with regard to the presence or absence of innermost gas and accreting modes, also covering both Group I and Group II SED shapes. Additional information, including Gaia EDR3 distances and self-consistently obtained stellar and circumstellar properties, has been compiled for these and for all HAeBes with similar interferometric results available in the literature. Based on the most complete dataset concerning K-band interferometric sizes, a reassessment of the size-luminosity relation has been made, leading to the following conclusions:
\begin{itemize}
    \item We confirm that the location of most HAes is consistent with the predictions of the optically-thin inner disk-MA model in the size-luminosity diagram, whereas a significant fraction of the more luminous HBes are "undersized" and fall within the range predicted by the classical-BL model with an optically thick gas disk that reaches the central star. 
    \item Based on the nUV Balmer excesses, the H$\alpha$ luminosities, the accretion luminosities, or the mass accretion rates, there is no evidence supporting the hypothesis that the shielding effect caused by innermost gas is the general reason explaining the comparatively smaller inner dust sizes of some HAeBes. Underlying relations between the three latter parameters and the stellar luminosity must be removed before any comparison is carried out. The few stars for which direct information on the presence of innermost atomic gas inferred from spectro-interferometry is available do not support the above mentioned hypothesis either.
    \item Similarly, the more recent suggestion that stars with Group I SEDs have inner dust sizes larger than Group II sources is not supported from our data for any luminosity bin. In fact, the opposite trend is observed for the luminous stars with L$_*$ $>$ 1000 L$_{\odot}$, although this sub-sample is biased towards Group I stars, and additional luminous Group II sources are necessary to test such a possible trend.
    \item Based on the stars for which multi-epoch interferometric data is available, the reported variations of the inner dust sizes cannot explain alone the observed scatter in the size-luminosity diagram ($\geq$ $\pm$0.5 dex for a given stellar luminosity).
    \item The size-luminosity correlation is statistically significant only for the sub-sample made of TTs and low-luminosity HAes located at $<$ 250 pc, but the overall correlation is influenced by a common dependence of both parameters compared on the distance to the sources. However, although future improvements in our observational capabilities will break the dependencies with the distance, the overall size-luminosity relation will most probably remain essentially the same. Thus, the currently observed size-luminosity correlation can in principle be interpreted physically in spite of the mentioned dependencies with the distance.
\end{itemize}

Our study is based on an  investigation of the main proposed scenarios potentially explaining the different observed positions of HAeBes in the size-luminosity diagram. No associated observational trend capable of confirming any such hypotheses has been found. Thus, a general rationale for the distribution of the stars in the size-luminosity correlation is still lacking.

%--------------------------------------------------------------------
%--------------------------------------------------------------------
%-----------------------------------------------------------------
\begin{acknowledgements}
Based on observations collected at the European Southern Observatory under ESO programme 0102.C-0576. The authors acknowledge the anonymous referee for the useful comments, which have served to improve the manuscript. PMA, IM, and JGD acknowledge the
Government of Comunidad Autónoma de Madrid (Spain) for funding this
research through a ‘Talento’ Fellowship (2016-T1/TIC-1890, PI I. Mendigut\'ia). The research of IM, JGD, and BM is also partially funded by the Spanish "Ministerio de Ciencia, Innovaci\'on y Universidades" through the national project "On the Rocks II" (PGC2018-101950-B-100; PI E. Villaver). 
EK is funded by the STFC (ST/P00041X/1). MV acknowledges the STARRY project, which has received funding from the European Union’s Horizon 2020 research and innovation programme under MSCA ITN-EID grant agreement No 676036. The research leading to these results has received funding from the European Union’s Horizon 2020 research and innovation programme under Grant Agreement 730890 (OPTICON).
This research has made use of the Jean-Marie Mariotti Center \texttt{LITpro}\footnote{LITpro software available at http://www.jmmc.fr/litpro} service co-developed by CRAL, IPAG and LAGRANGE. PMA acknowledges M. Tallon in particular for his support on LITpro, and A.C. Carciofi and R. G. Vieira for their contribution to the observing proposal leading to these results.

\end{acknowledgements}

   \bibliographystyle{aa} % style aa.bst
   \bibliography{marcos_2020} % your references Yourfile.bib
\begin{appendix} %First appendix
\section{Log of the observations and interferometric results}
\label{Appendix:observations}
\begin{figure}
   \centering
   \includegraphics[width=9cm]{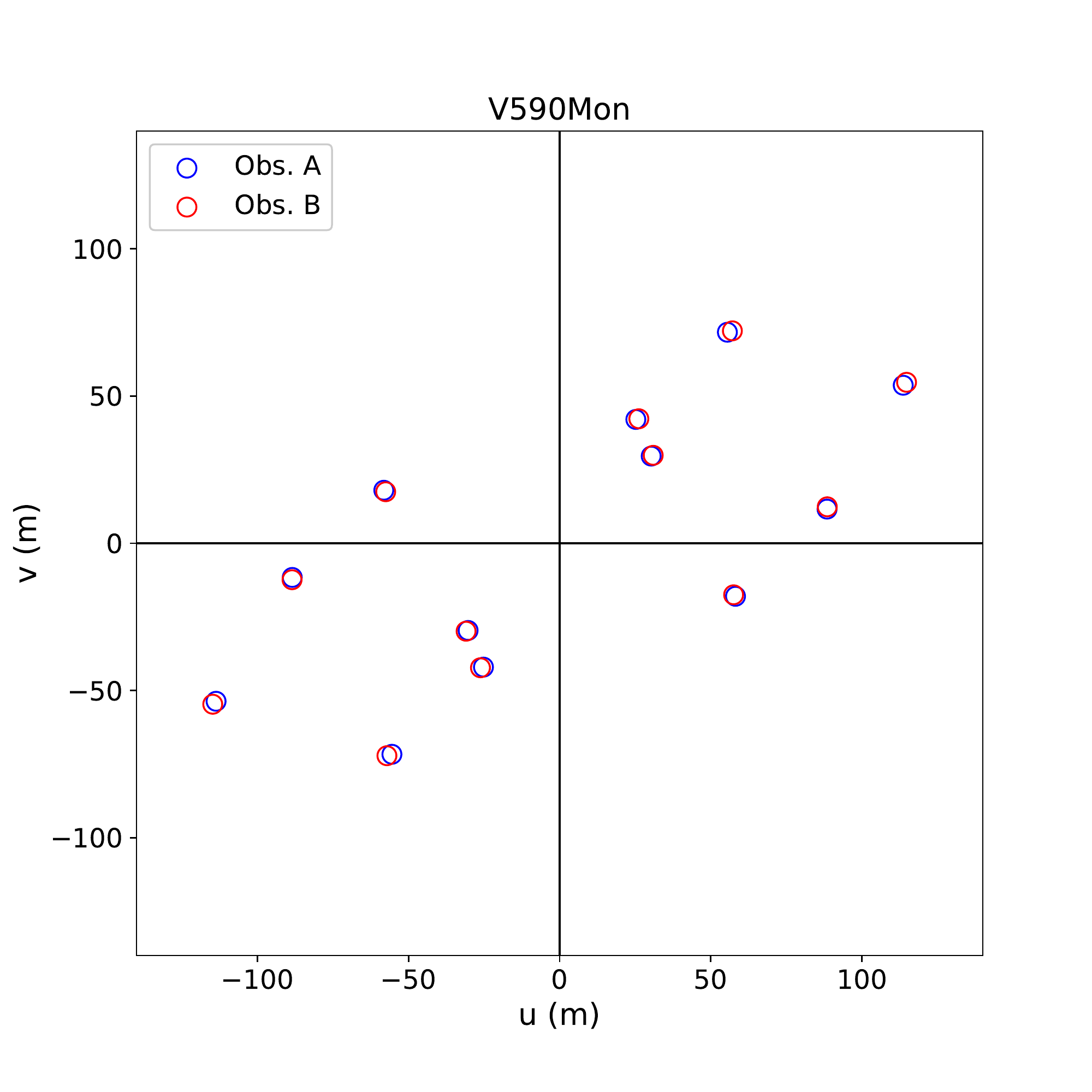}
      \caption{uv coverage of the two V590 Mon observations (noted as Obs. A and Obs. B).
      }
         \label{fig:V590Mon_u_v_coverage}
   \end{figure}

\begin{figure}
   \centering
   \includegraphics[width=9cm]{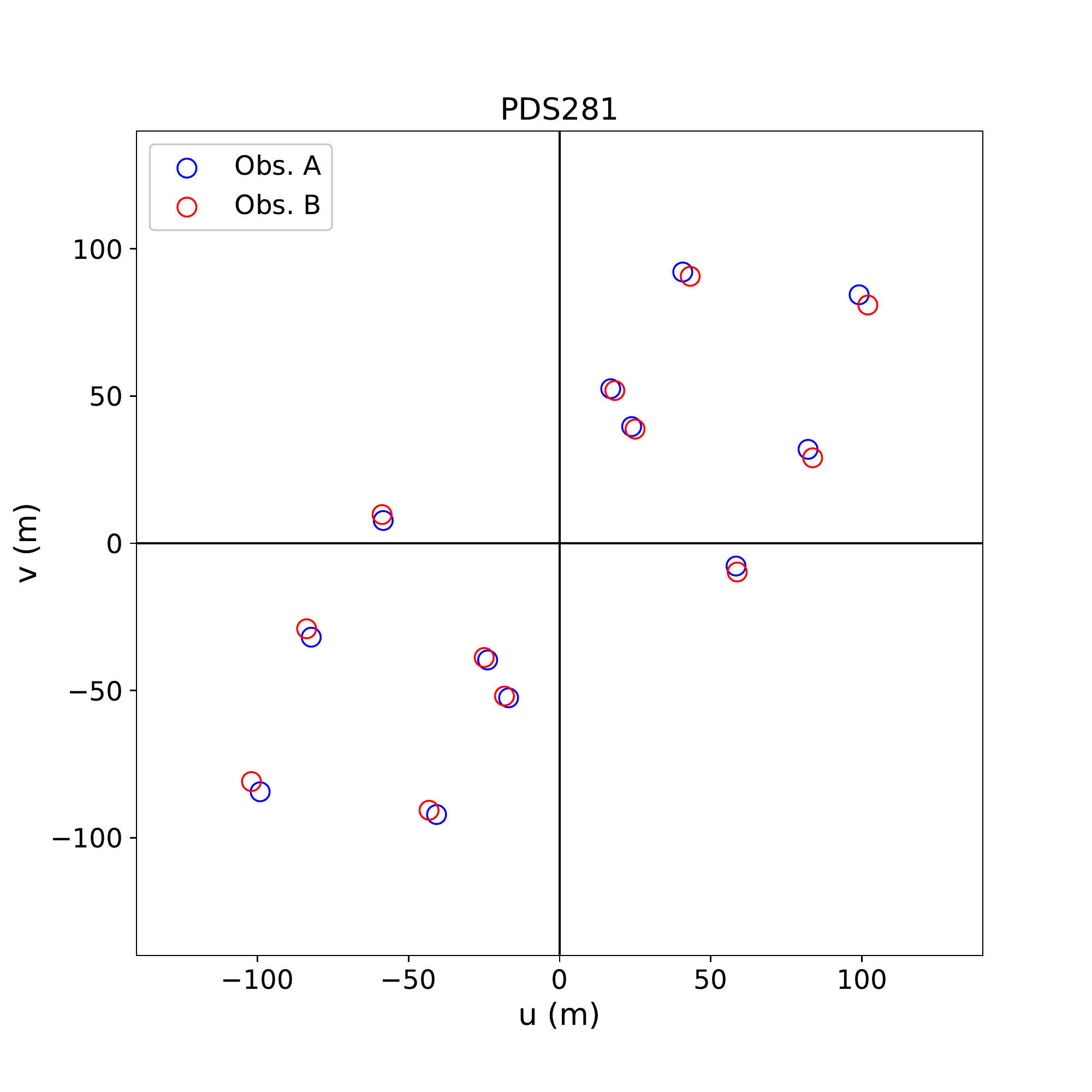}
      \caption{uv coverage of the two PDS 281 observations (noted as Obs. A and Obs. B).
      }
         \label{fig:PDS281_u_v_coverage}
   \end{figure}

\begin{figure}
   \centering
   \includegraphics[width=9cm]{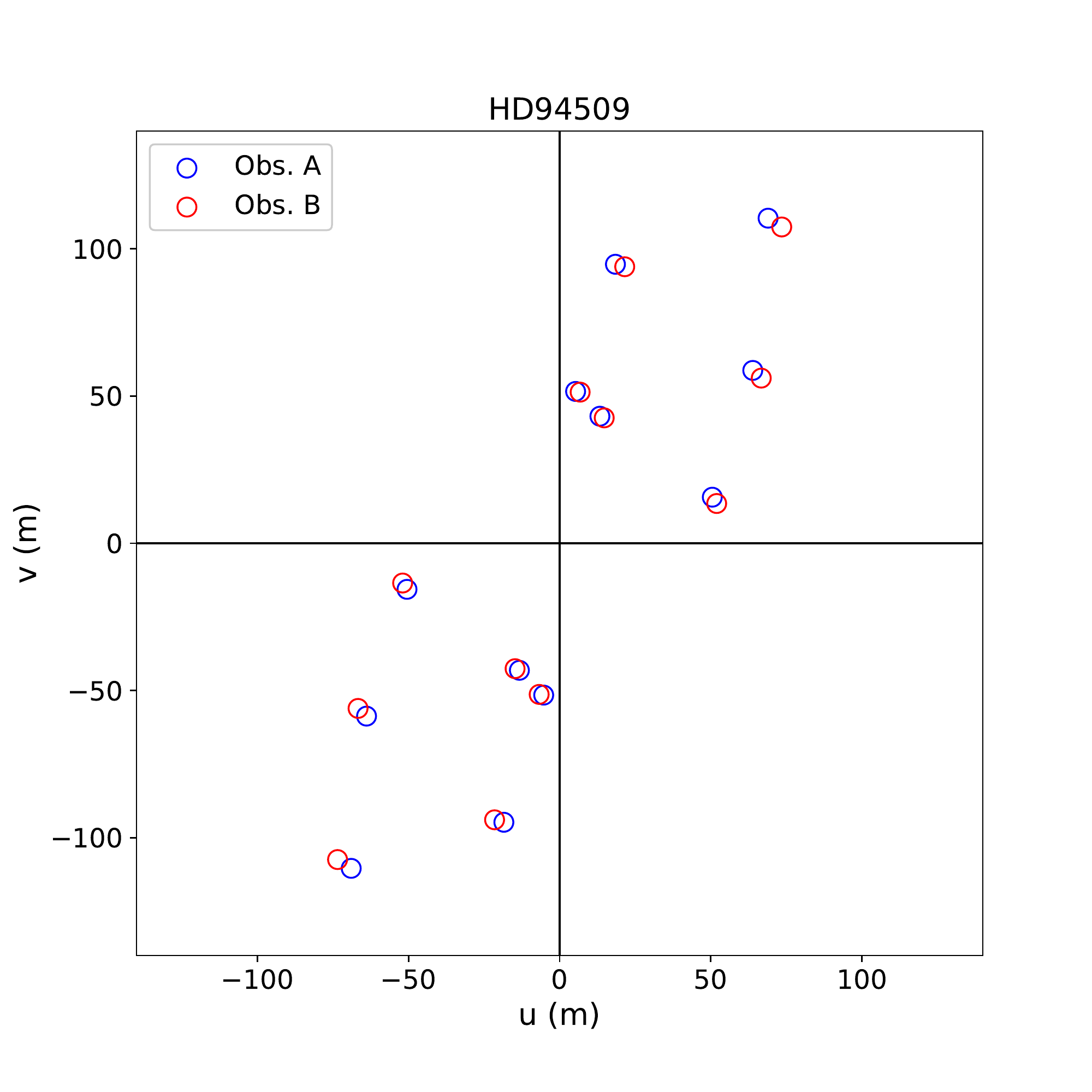}
      \caption{uv coverage of the two HD 94509 observations (noted as Obs. A and Obs. B).
      }
         \label{fig:HD94509_u_v_coverage}
   \end{figure}

\begin{figure}
   \centering
   \includegraphics[width=9cm]{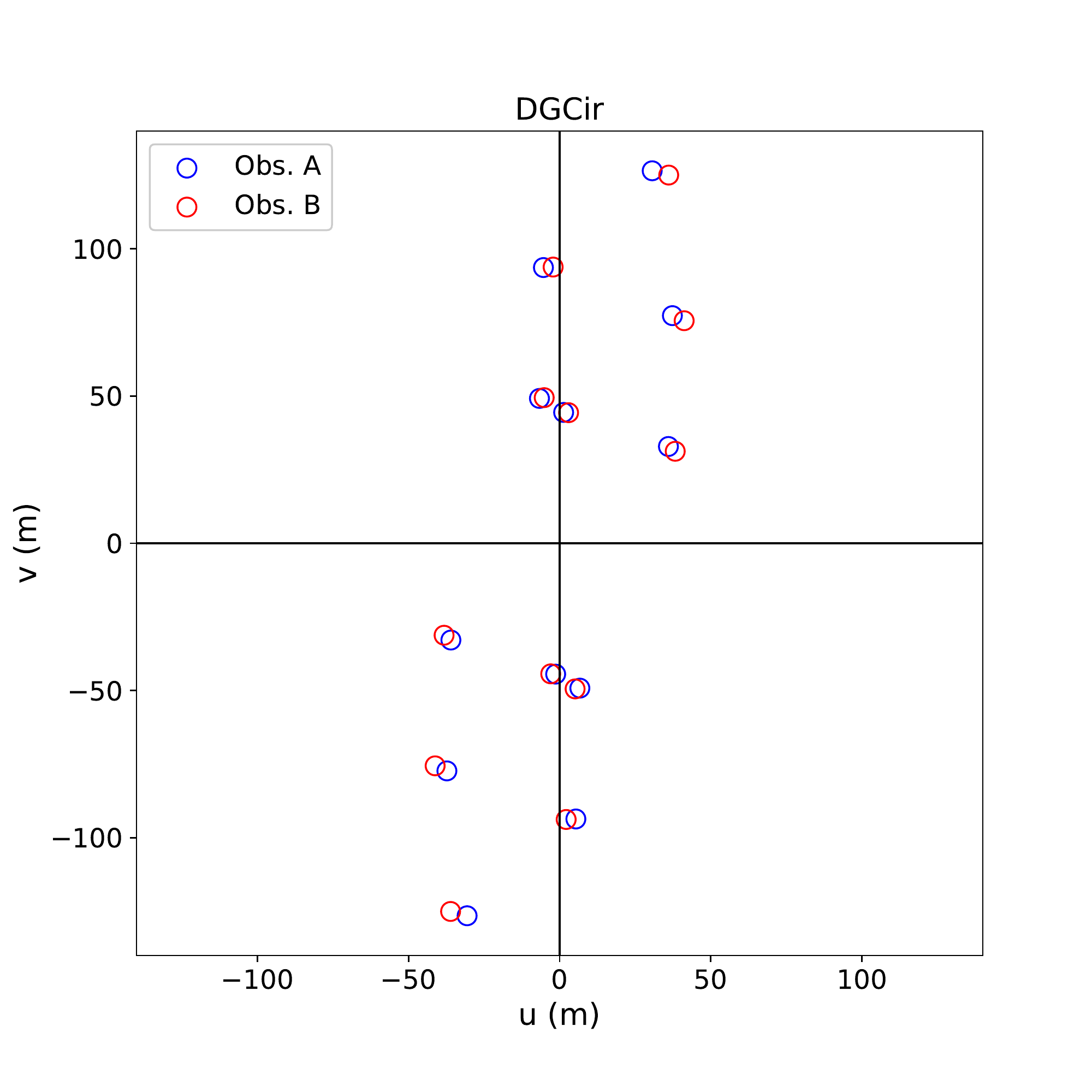}
      \caption{uv coverage of the two DGCir observations (noted as Obs. A and Obs. B).
      }
         \label{fig:DGCir_u_v_coverage}
   \end{figure}

\begin{figure}
   \centering
   \includegraphics[width=9cm]{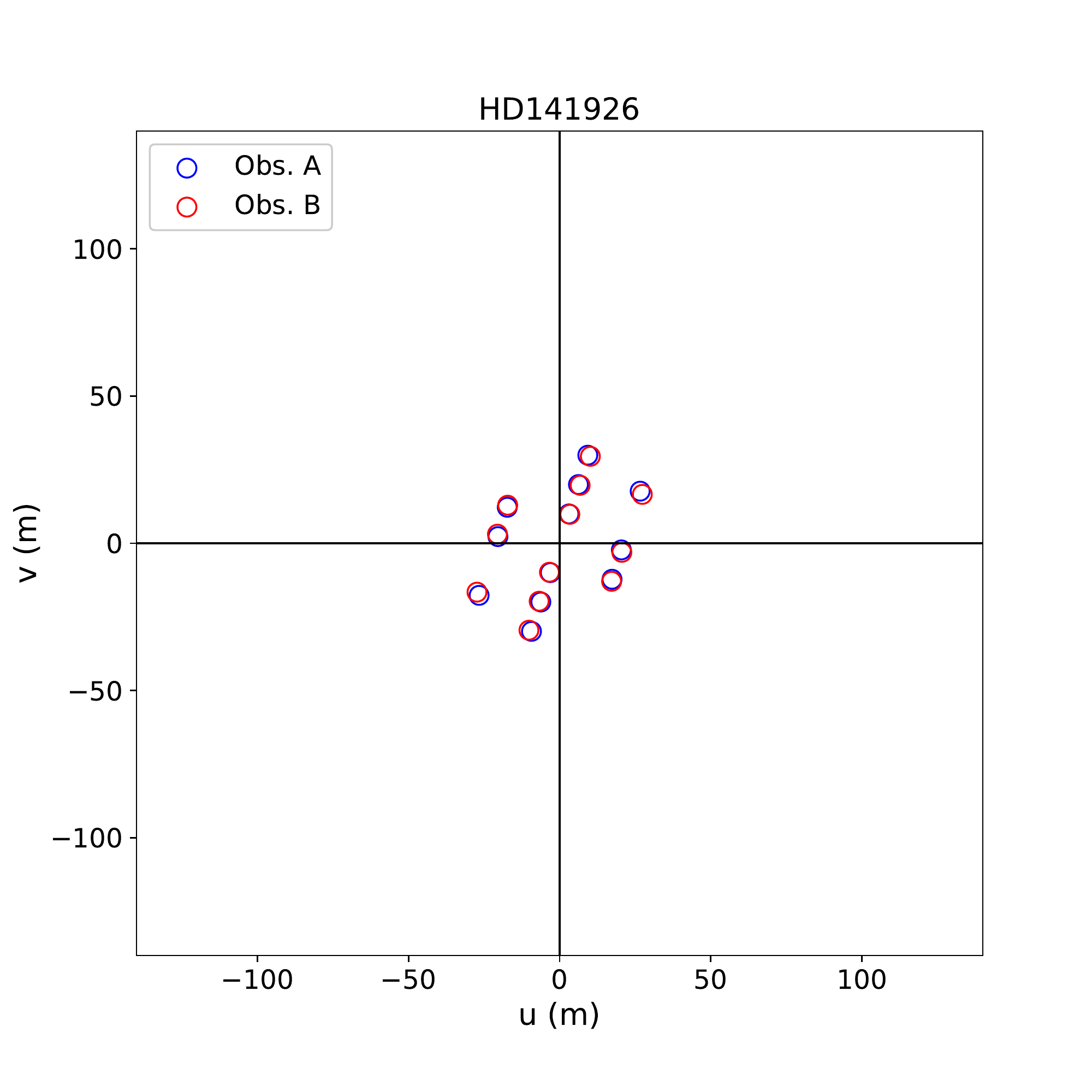}
      \caption{uv coverage of the two HD 141926 observations (noted as Obs. A and Obs. B). This is the only star for which the 1.8m ATs were used instead of the 8.2m UTs (Sect. \ref{sec:obs}).
      }
         \label{fig:HD141926_u_v_coverage}
   \end{figure}

\begin{figure}
   \centering
   \includegraphics[width=9cm,trim=25 55 40 70, clip]{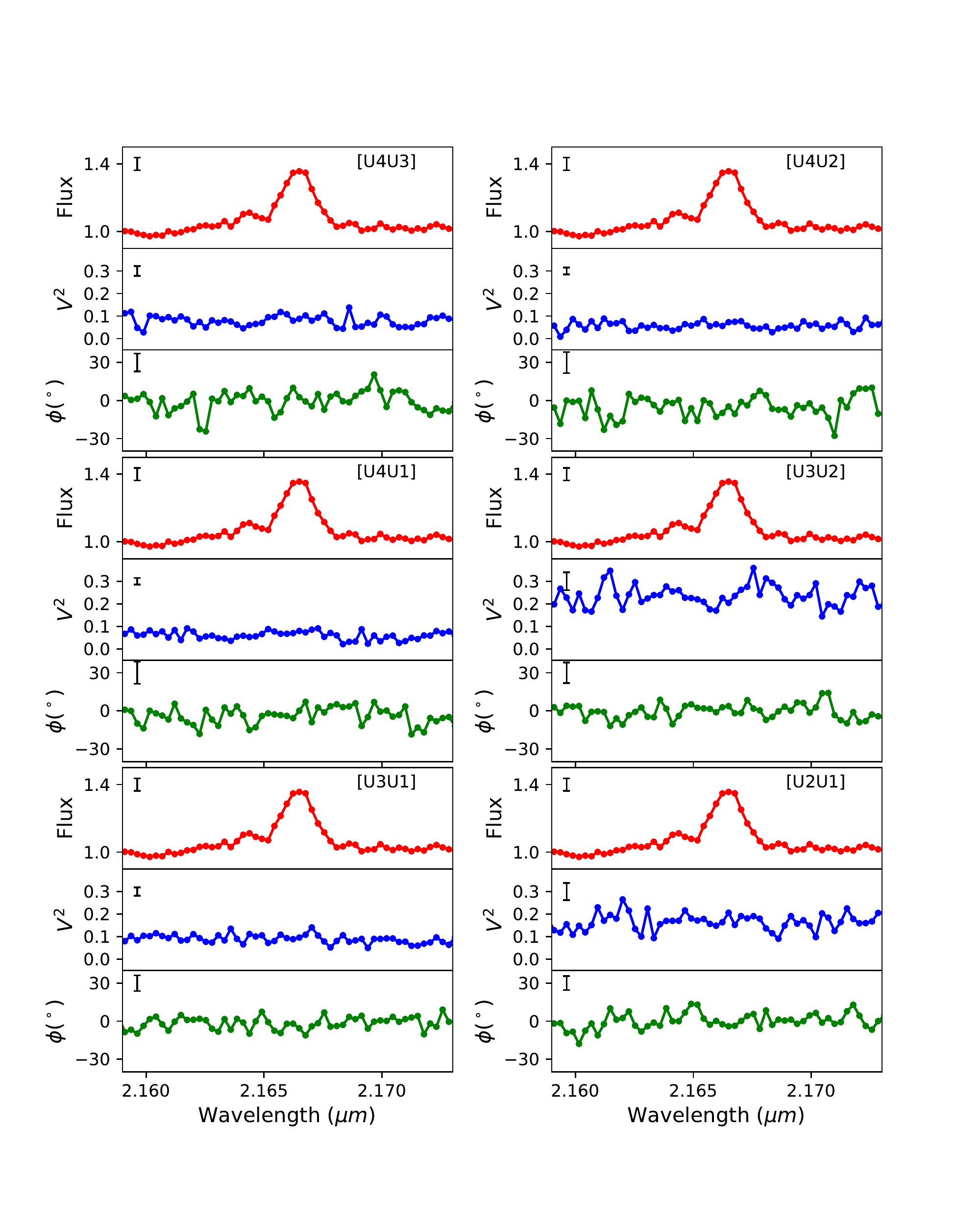}
      \caption{Composition of six triplets of panels corresponding to each V590 Mon observation baseline. For each baseline is shown the normalized flux (in red; the same for all baselines), squared visibilities (in blue) and differential phases (in green) in a wavelength range centered in Br$\gamma$. Typical uncertainties are indicated in each panel, representing the standard deviation of the corresponding parameter in the adjacent continuum of the line. 
      }
         \label{fig:V590Mon_3plots}
   \end{figure}

\begin{figure}
   \centering
   \includegraphics[width=9cm,trim=25 55 40 70, clip]{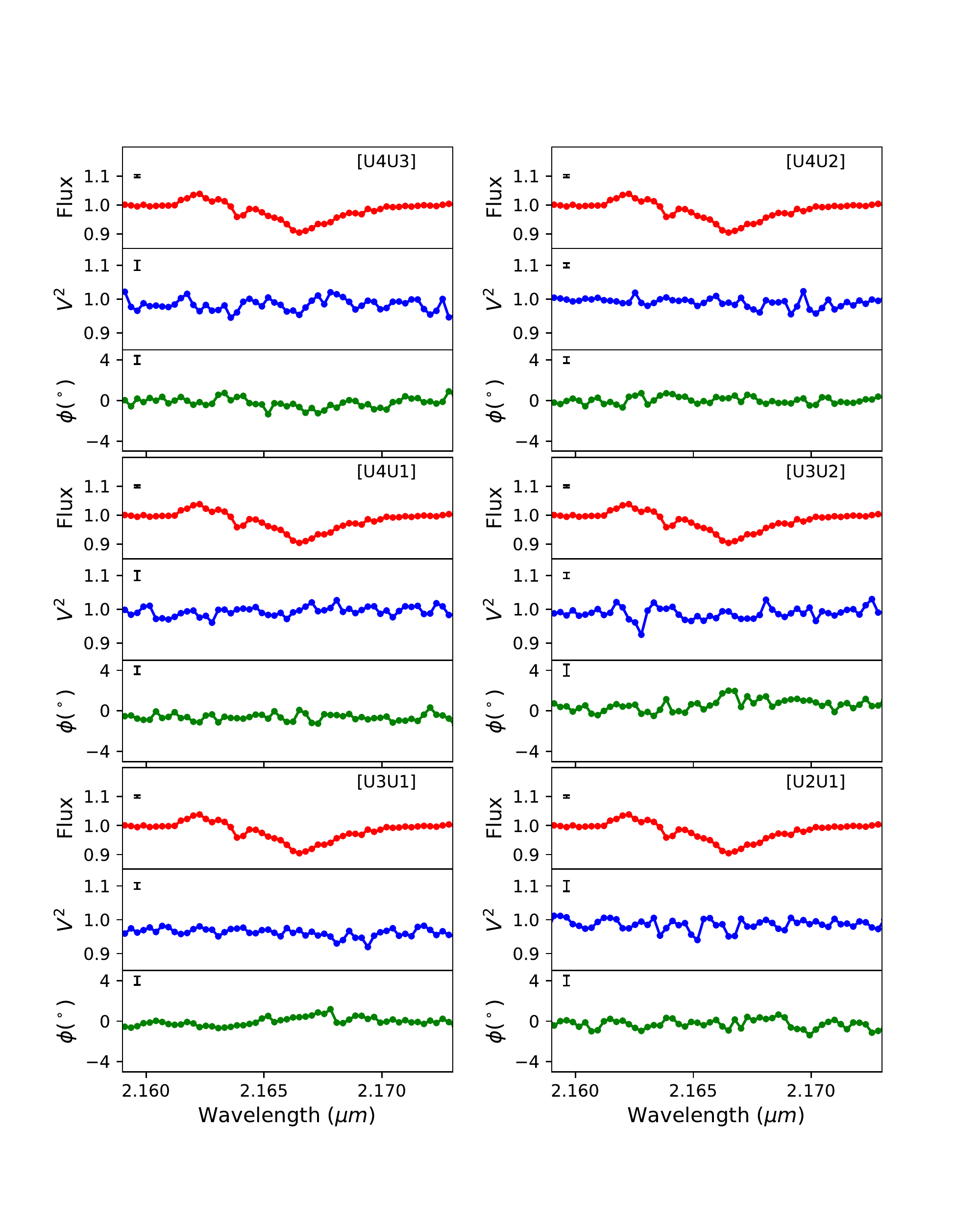}
      \caption{Composition of six triplets of panels corresponding to each PDS 281 observation baseline (see the caption of Fig. \ref{fig:V590Mon_3plots}) 
      }
         \label{fig:PDS281_3plots}
   \end{figure}

\begin{figure}
   \centering
   \includegraphics[width=9cm,trim=25 55 40 70, clip]{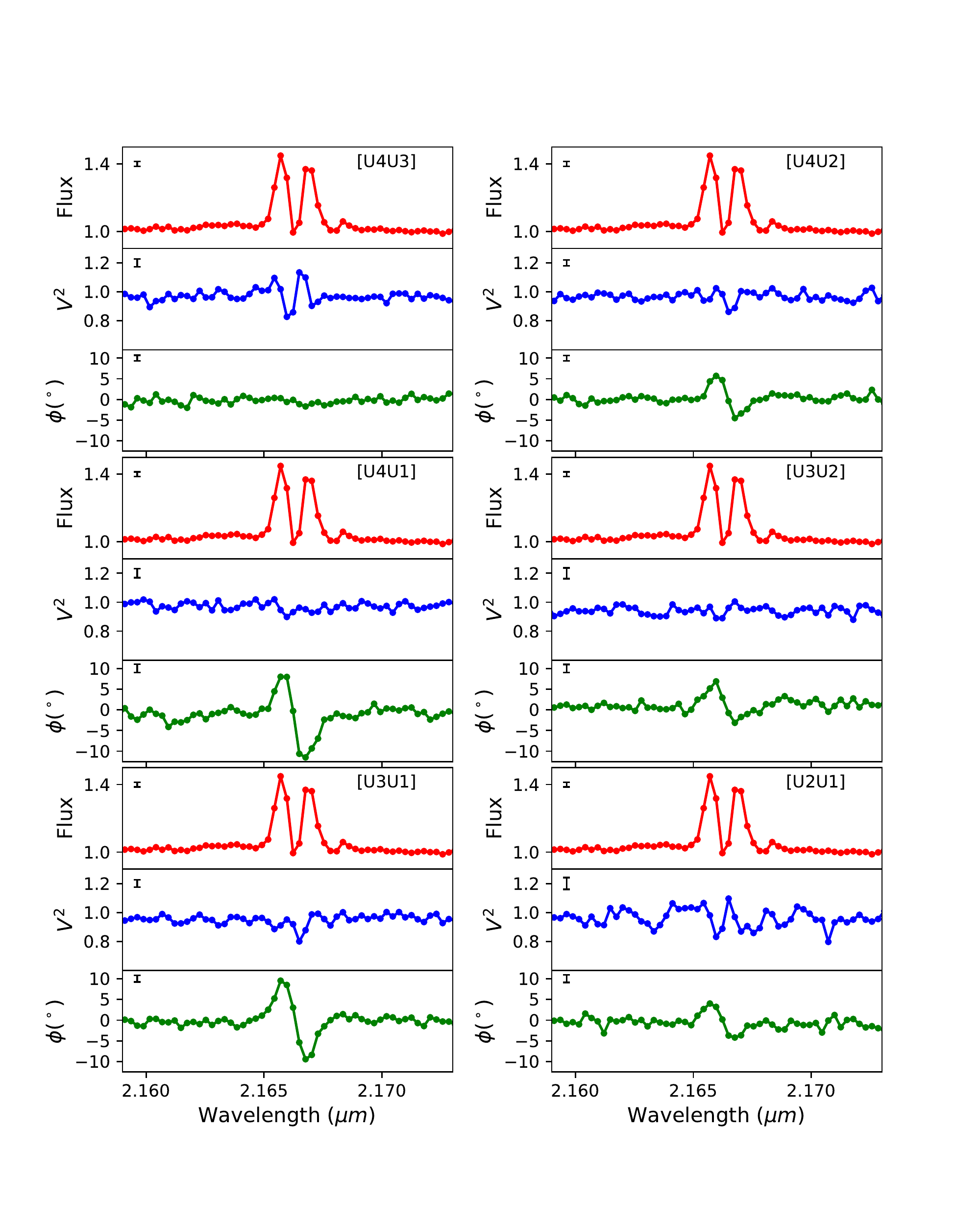}
      \caption{Composition of six triplets of panels corresponding to each HD 94509 observation baseline (see the caption of Fig. \ref{fig:V590Mon_3plots}) 
      }
         \label{fig:HD94509_3plots}
   \end{figure}

\begin{figure}
   \centering
   \includegraphics[width=9cm,trim=25 55 40 70, clip]{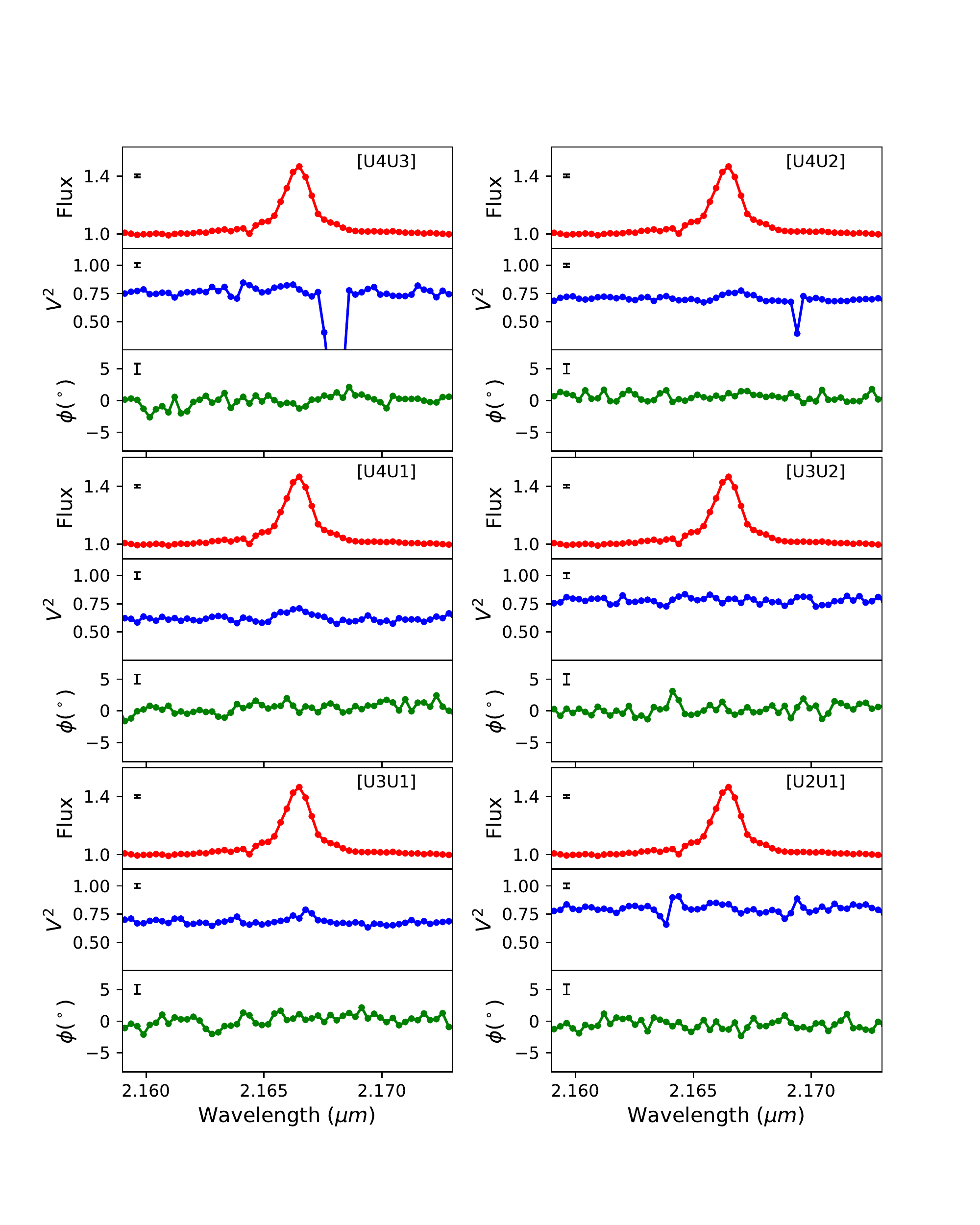}
      \caption{Composition of six triplets of panels corresponding to each DG Cir observation baseline (see the caption of Fig. \ref{fig:V590Mon_3plots}) 
      }
         \label{fig:DGCir_3plots}
   \end{figure}

\begin{figure}
   \centering
   \includegraphics[width=9cm,trim=25 55 40 70, clip]{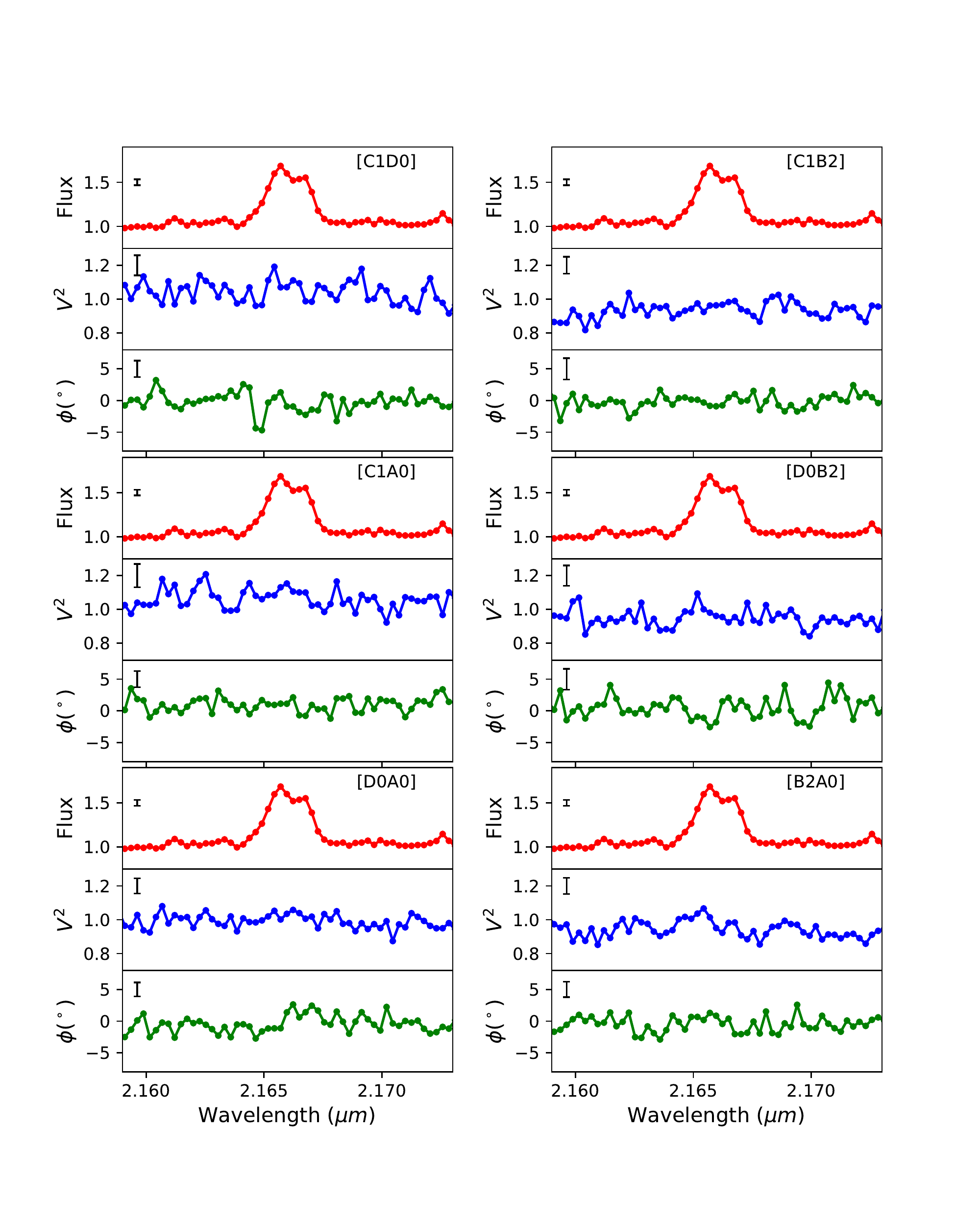}
      \caption{Composition of six triplets of panels corresponding to each  HD 141926 observation baseline (see the caption of Fig. \ref{fig:V590Mon_3plots})  
      }
         \label{fig:HD141926_3plots}
   \end{figure}

\begin{figure*}
   \centering
   \includegraphics[width=18cm]{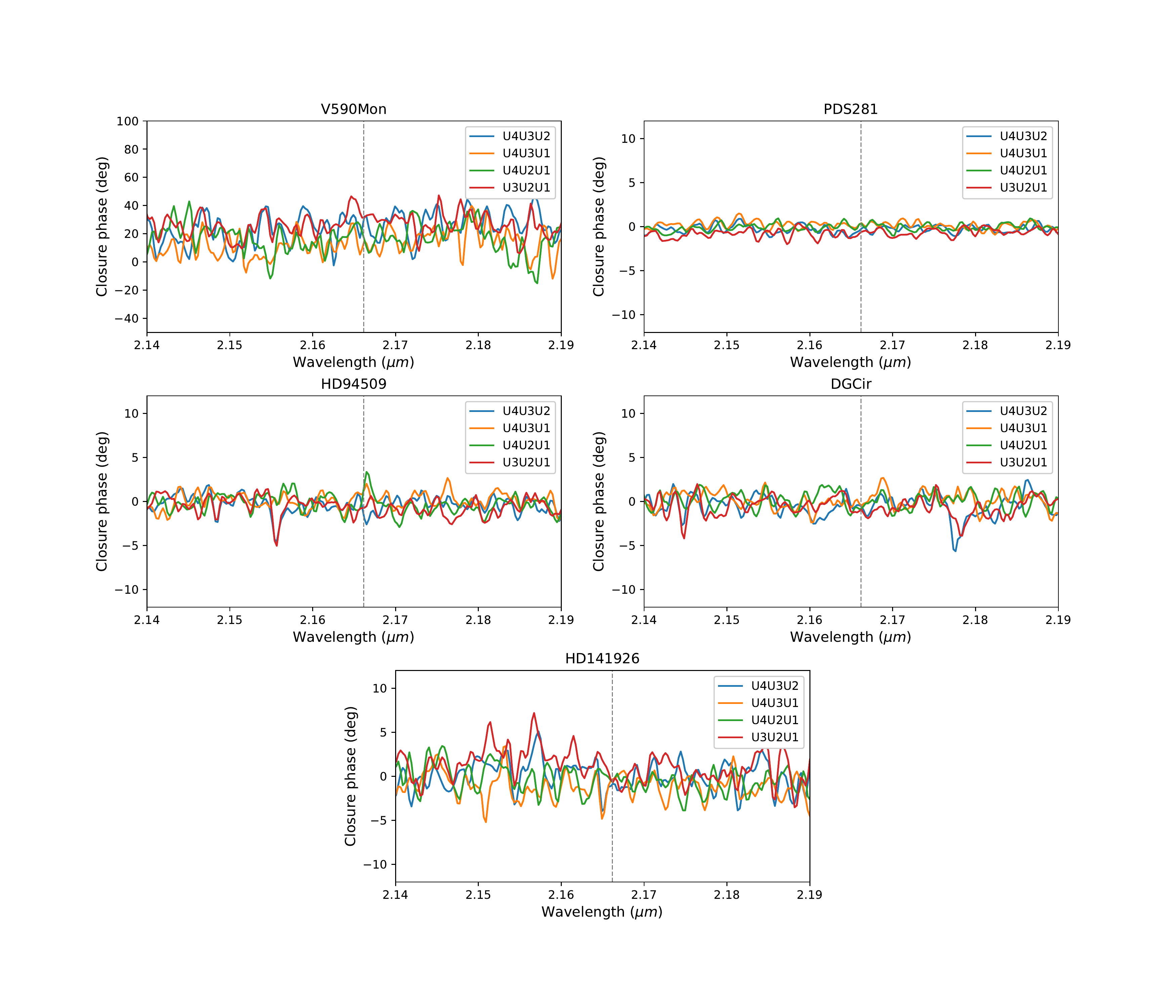}
      \caption{Closure phases of the stars in our sample around Br$\gamma$ line, which center is indicated with the vertical line. Different colors indicate different triplets, as indicated. 
      }
         \label{fig:closure_phases_Brg}
   \end{figure*}

Table \ref{table:A1} includes the target stars, their coordinates and K-magnitudes, the corresponding calibrators and magnitudes, observing dates, average values of the seeing, coherence times, and the telescopes used. 
Figures \ref{fig:V590Mon_u_v_coverage}, \ref{fig:PDS281_u_v_coverage}, \ref{fig:HD94509_u_v_coverage}, \ref{fig:DGCir_u_v_coverage}, and \ref{fig:HD141926_u_v_coverage} show the uv coverage for each target star; Figs. \ref{fig:V590Mon_3plots}, \ref{fig:PDS281_3plots}, \ref{fig:HD94509_3plots}, \ref{fig:DGCir_3plots}, and \ref{fig:HD141926_3plots} their observed fluxes, squared visibilities, and differential phases from top to bottom for the different baselines. For Fig. \ref{fig:closure_phases_Brg}, the closure phases per star and baseline are given. 

\begin{table*}
\caption{Log of the observations}
\label{table:A1}
\centering
\renewcommand\tabcolsep{2.5pt}
\begin{tabular}{l c c c c c c c c c}
\hline\hline
   Object &          RA &        DEC &    $K_{Object}$& Calibrator &$K_{Calibrator}$& Observation&  Seeing &Coherence  & Telescopes \\
          &     [h:m:s] &   [deg:m:s]&  [mag]         &            &  [mag]         &     date   &         & Time\tablefootmark{a} [ms]       &            \\
%          &     (h:m:s) & \multicolumn{7}{l}{(deg:m:s)} \\
\hline   
  V590Mon &  06:40:44.6 &  +09:48:02 & $9.33\pm 0.03$ &   HD259163 & $9.26\pm0.02$  & 27-10-2018 &  0.42'' & 4, 6   & U1-U2-U3-U4 \\
   PDS281 &  08:55:45.9 &  -44:25:14 &$7.32\pm0.02$   &    HD78958 & $7.22\pm0.03$  & 20-11-2018 &  0.57'' & 3, 4   & U1-U2-U3-U4 \\
  HD94509 &  10:53:27.2 &  -58:25:24 & $8.93\pm 0.02$ &    HD94533 & $8.94\pm0.02$  & 20-12-2018 &  0.36'' & 9, 10  & U1-U2-U3-U4 \\
    DGCir &  15:03:23.8 &  -63:22:59 &$7.82\pm0.02$   &   HD131662 & $7.77\pm0.02$  & 22-01-2019 &  0.46'' & 17, 18 & U1-U2-U3-U4 \\
 HD141926 &  15:54:21.8 &  -55:19:44 & $6.51\pm 0.02$ &   HD145664 & $7.44\pm0.04$  & 21-03-2019 &  0.45'' & 7, 8   & A0-B2-C1-D0 \\
\hline
\end{tabular}
\tablefoot{
\tablefoottext{a}{Each object was observed twice. The exposure time for all observations of all objects were 30 s.}
}
\end{table*}

%-------------------------------------------------------------
%                                Figures/V590Mon_u_v_coverage
%-------------------------------------------------------------
%-------------------------------------------------------------
%-------------------------------------------------------------
%                                Figures/PDS281_u_v_coverage.eps 
%-------------------------------------------------------------
%--------------------------------------------------------------------
%-------------------------------------------------------------
%                                Figures/HD94509_u_v_coverage 
%-------------------------------------------------------------
%--------------------------------------------------------------------

%-------------------------------------------------------------
%                                Figures/DGCir_u_v_coverage 
%-------------------------------------------------------------
%--------------------------------------------------------------------

%-------------------------------------------------------------
%                                Figures/HD141926_u_v_coverage.eps 
%-------------------------------------------------------------

%-------------------------------------------------------------

%-------------------------------------------------------------
%                                Figures/V590Mon_figure_3_plots
%-------------------------------------------------------------
%-------------------------------------------------------------
%-------------------------------------------------------------
%                                Figures/PDS281_figure_3_plots.eps 
%-------------------------------------------------------------
%--------------------------------------------------------------------
%-------------------------------------------------------------
%                                Figures/HD94509_figure_3_plots 
%-------------------------------------------------------------
%--------------------------------------------------------------------

%-------------------------------------------------------------
%                                Figures/DGCir_figure_3_plots 
%-------------------------------------------------------------
%--------------------------------------------------------------------

%-------------------------------------------------------------
%                                Figures/HD141926_figure_3_plots.eps 
%-------------------------------------------------------------

%-------------------------------------------------------------
%                                Figures/closure_phases_Brg.eps 
%-------------------------------------------------------------

\end{appendix}

\end{document}